\documentclass[preprint,twocolumn]{aastex631}
\usepackage[utf8]{inputenc}
\usepackage{savesym}
\savesymbol{tablenum}
\usepackage{siunitx}
\restoresymbol{SIX}{tablenum}
\usepackage{amsmath}
\usepackage{natbib}
\usepackage{float}
\usepackage{appendix}
\usepackage{multirow}
\usepackage{booktabs}
\usepackage{hyperref}
\usepackage{comment}
\newcommand{\CellWithForceBreak}[2][c]{\begin{tabular}[#1]{@{}c@{}}#2\end{tabular}}

\begin{document}

\title{JWST-TST High Contrast: First Direct Spectroscopy of GJ 504 b reveals Clouds and Possible Metal Enrichment}

\author[0000-0003-3708-241X]{Aneesh Baburaj} \affiliation{Department of Astronomy \& Astrophysics, University of California, San Diego, La Jolla, CA 92093, USA} \affiliation{Department of Physics, University of California, San Diego, La Jolla, CA 92093, USA}\affiliation{Center for Interdisciplinary Exploration and Research in Astrophysics, Northwestern University, 1800 Sherman Ave, Evanston, IL 60201, USA}

\author[0000-0003-2233-4821]{Jean-Baptiste Ruffio} \affiliation{Department of Astronomy \& Astrophysics, University of California, San Diego, La Jolla, CA 92093, USA}

\author[0000-0002-3191-8151]{Marshall Perrin} \affiliation{Space Telescope Science Institute, 3700 San Martin Drive, Baltimore, MD 21218, USA}

\author[0000-0002-6618-1137]{Jerry W. Xuan}
\altaffiliation{51 Pegasi b Fellow}
\affiliation{Department of Earth, Planetary, and Space Sciences, University of California, Los Angeles, CA 90095, USA}
\affiliation{Department of Astronomy, California Institute of Technology, Pasadena, CA 91125, USA}

\author[0000-0001-6396-8439]{William O. Balmer} \affiliation{William H. Miller III Department of Physics and Astronomy, Johns Hopkins University, Baltimore, MD 21218, USA} \affiliation{Space Telescope Science Institute, 3700 San Martin Drive, Baltimore, MD 21218, USA}

\author[0000-0003-1728-8269]{Yayaati Chachan} \affiliation{Department of Astronomy \& Astrophysics, University of California, Santa Cruz, CA95064, USA}

\author[0000-0002-9936-6285]{Quinn M. Konopacky} \affiliation{Department of Astronomy \& Astrophysics, University of California, San Diego, La Jolla, CA 92093, USA} 

\author[0000-0002-7129-3002]{Travis S. Barman} \affiliation{Lunar and Planetary Laboratory, University of Arizona, Tucson, AZ 85721, USA}

\author[0000-0002-2918-8479]{Mathilde M\^alin} \affiliation{William H. Miller III Department of Physics and Astronomy, Johns Hopkins University, Baltimore, MD 21218, USA} \affiliation{Space Telescope Science Institute, 3700 San Martin Drive, Baltimore, MD 21218, USA}

\author[0000-0002-9803-8255]{Kielan K. W. Hoch} \affiliation{Space Telescope Science Institute, 3700 San Martin Drive, Baltimore, MD 21218, USA}

\author[0000-0003-4203-9715]{Emily Rickman} \affiliation{European Space Agency (ESA), ESA Office, Space Telescope Science Institute, 3700 San Martin Dr, Baltimore, MD 21218, USA}

\author[0000-0002-4479-8291]{Kimberly Ward-Duong} \affiliation{Smith College, 1 Chapin Way, Northampton, MA 01063, USA}

\author[0000-0003-3818-408X]{Laurent Pueyo} \affiliation{Space Telescope Science Institute, 3700 San Martin Drive, Baltimore, MD 21218, USA}

\author[0000-0001-8627-0404]{Julien H. Girard} \affiliation{Space Telescope Science Institute, 3700 San Martin Drive, Baltimore, MD 21218, USA}

\author[0000-0002-4388-6417]{Isabel Rebollido} \affiliation{European Space Agency (ESA), European Space Astronomy Centre (ESAC), Camino Bajo del Castillo s/n, 28692 Villanueva de la Ca\~nada, Madrid, Spain}

\author[0000-0002-9799-2303]{Alexis Bidot} \affiliation{Space Telescope Science Institute, 3700 San Martin Drive, Baltimore, MD 21218, USA}

\author[0000-0002-8382-0447]{Christine Chen} \affiliation{Space Telescope Science Institute, 3700 San Martin Drive, Baltimore, MD 21218, USA}

\author[0000-0002-5885-5779]{Kadin Worthen} \affiliation{William H. Miller III Department of Physics and Astronomy, Johns Hopkins University, Baltimore, MD 21218, USA}

\author[0000-0001-9352-0248]{Cicero Lu} \affiliation{Gemini Observatory/NSF NOIRLab, 670 N. A’ohoku Place, Hilo, Hawai’i, 96720, USA}

\author[0000-0003-2769-0438]{Jens Kammerer} \affiliation{European Southern Observatory, Karl-Schwarzschild-Str. 2, 85748 Garching, Germany}

\author[0000-0001-7827-7825]{Roeland P. van der Marel} \affiliation{Space Telescope Science Institute, 3700 San Martin Drive, Baltimore, MD 21218, USA} \affiliation{William H. Miller III Department of Physics and Astronomy, Johns Hopkins University, Baltimore, MD 21218, USA}

\author[0000-0002-8507-1304]{Nikole K. Lewis} \affiliation{Department of Astronomy and Carl Sagan Institute, Cornell University, 122 Sciences Drive, Ithaca, NY 14853, USA}

\author[0000-0003-3305-6281]{Jeff Valenti} \affiliation{Space Telescope Science Institute, 3700 San Martin Drive, Baltimore, MD 21218, USA}

\author[0000-0002-6892-6948]{Sara Seager} \affiliation{Department of Physics and Kavli Institute for Astrophysics and Space Research, Massachusetts Institute of Technology, Cambridge, MA 02139, USA} \affiliation{Department of Earth, Atmospheric, and Planetary Sciences, Massachusetts Institute of Technology, Cambridge, MA 02139, US} \affiliation{Department of Aeronautics and Astronautics, MIT, 77 Massachusetts Avenue, Cambridge, MA 02139, USA}

\author[0009-0003-0199-2822]{Chris Stark} \affiliation{NASA Goddard Space Flight Center, 8800 Greenbelt Road, Greenbelt, MD 20771, USA}

\author[0000-0003-2753-2819]{R\'emi Soummer} \affiliation{Space Telescope Science Institute, 3700 San Martin Drive, Baltimore, MD 21218, USA}

\author{Jay Anderson} \affiliation{Space Telescope Science Institute, 3700 San Martin Drive, Baltimore, MD 21218, USA}

\author[0009-0003-3993-8338]{Charles-Philippe Lajoie} \affiliation{Space Telescope Science Institute, 3700 San Martin Drive, Baltimore, MD 21218, USA}

\author[0000-0003-4003-8348]{Mark Clampin} \affiliation{Astrophysics Division, Science Mission Directorate, NASA Headquarters, 300 E Street SW, Washington, DC 20546, USA}

\author{C. Matt Mountain} \affiliation{Association of Universities for Research in Astronomy, 1331 Pennsylvania Avenue NW Suite 1475, Washington, DC 20004, USA}

\accepted{May 4, 2026}

\correspondingauthor{Aneesh Baburaj}
\email{ababuraj@northwestern.edu}

\begin{abstract}
Characterizing the coldest directly imaged companions through direct spectroscopy has only recently become possible with the James Webb Space Telescope. We present moderate-resolution (R $\sim$ 2,700) spectroscopic observations of the directly imaged planetary-mass companion (PMC), GJ 504 b, using the $JWST$/NIRSpec. As the coldest imaged PMC of the pre-JWST era GJ 504 b is too faint for ground-based spectroscopy, with only photometric observations possible. Leveraging advanced post-processing techniques with a forward modeling framework, we detect the companion at high signal-to-noise (S/N$>$300). We also present the first successful PSF subtraction with angular differential imaging (ADI) in the NIRSpec point cloud, detecting GJ 504 b at S/N$>10$ and reaching contrast limits $<10^{-4}$. The extracted 2.9--5.3 $\mu m$ spectra show strong signatures of several molecular species, including H$_2$O, $^{12}$C$^{16}$O, CH$_4$, CO$_2$, NH$_3$, H$_2$S, $^{13}$C$^{16}$O, and $^{12}$C$^{18}$O. Atmospheric modeling of the spectra using \texttt{petitRADTRANS}, yields an effective temperature = 564$\pm$4 K, surface gravity $\log{g}$ = 4.87$^{+0.13}_{-0.12}$, metallicity [M/H] = 0.67$^{+0.13}_{-0.12}$, C/O ratio = 0.64$^{+0.02}_{-0.02}$, interstellar $^{12}$C/$^{13}$C and $^{16}$O/$^{18}$O isotopologue ratios, and strong evidence of disequilibrium chemistry and salt clouds. The retrieved parameters indicate a mass 25.2$^{+8.4}_{-6.0}$ $M_\mathrm{Jup}$, which is in agreement with the mass range (19--27 $M_\mathrm{Jup}$) obtained from ATMO evolutionary models, implying an age of 2.5--4.0 Gyr. Lastly, we compare the abundances of GJ 504 b to its primary, obtaining a stellar abundance of sulfur (S), super-stellar carbon (C), and possibly, oxygen (O). The observed metal enrichment tentatively supports planet-like formation, but does not entirely exclude stellar abundances for GJ 504 b.
\end{abstract}

\keywords{Direct imaging; Exoplanet atmospheric composition; Atmospheric clouds; Exoplanet formation}

\section{Introduction}
Direct spectroscopy is a powerful technique that enables investigation of atmospheric properties of an exoplanet independent of the properties of the host star. In particular, spectral characterization using moderate-to-high resolution spectroscopy can provide information about effective temperature, surface gravity, and metallicity, as well as individual elemental abundances and cloud properties (e.g., \citealt{2013Sci...339.1398K, 2021AJ....162..290R, 2021Natur.595..370Z, xuan2022, 2023ApJ...946L...6M}). 
Such detailed atmospheric characterization is especially relevant in the case of the super-Jovian, wide orbit directly imaged exoplanets (sma $>5\,$au, $M_p>2\,M_{\rm Jup}$), whose formation has proved challenging to explain by the classical core accretion and gravitational instability pathways \citep[e.g.,][]{2009ApJ...707...79D,2016ARA&A..54..271K}. The recently revised mass-metallicity relations indicate that core accretion can form planets beyond the deuterium burning limit, with emerging evidence indicating that even super-Jupiters at tens of au are metal-rich and distinct from brown-dwarf like objects (which tend to cluster around solar metallicity) \citep{2025ApJ...994...43C}. Initially, the atmospheric carbon-to-oxygen ratio (C/O) of exoplanets was proposed as a potential formation tracer  \citep{oberg2011, madhu2011, 2019ARA&A..57..617M}, prompting the use of molecular features from species like H$_2$O and CO to measure the C/O ratio and C/H and deduce formation pathways \citep[e.g.][]{2013Sci...339.1398K,2020AJ....160..207W,2021A&A...648A..59P,2021AJ....162..290R,2023AJ....166...85H, xuan2024b, balmer2025}. However, the degeneracies in formation histories from using just this formation diagnostic (e.g., \citealt{sb2021b, 2022ApJ...934...74M}) have prompted theoretical investigations into species like nitrogen \citep{2021ApJ...909...40T,2022ApJ...937...36P,2023ApJ...946...18O}, and refractory elements (sulfur, sodium, potassium, and silicon; \citealt{sb2021b, 2023ApJ...952L..18C, 2023ApJ...943..112C}), that would enable us to deduce formation mechanisms and locations more conclusively. The measurement of multiple elemental abundances would also enable a more robust estimate of the planetary metallicity.

\par Ground-based medium-resolution spectrographs like Keck/OSIRIS \citep{2006SPIE.6269E..1AL} and VLT/SINFONI \citep{2003SPIE.4841.1548E,2004Msngr.117...17B} can only provide spectral data over the 1--2.5 $\mu m$ range. With $JWST$ providing access to medium-resolution spectroscopy ($R$ $\sim$ 2,700) beyond 2.5 $\mu m$, molecular features due to CO$_2$, CH$_4$, NH$_3$, H$_2$S, among others, become accessible \citep{2022A&A...658A..72P,2023A&A...671A.109M, 2023ApJ...946L...6M,Ruffio2026}. NH$_3$ and H$_2$S allow us to utilize nitrogen and refractory-based planet formation tracers for formation deductions. Additionally, the wavelength coverage of $JWST$ over the near-infrared and mid-infrared coupled with its exquisite sensitivity has enabled the discovery and characterization of colder planets ($T_\mathrm{eff} <$ 500 K) than before (e.g., eps Ind b: \citealt{2024Natur.633..789M}, TWA 7 b: \citealt{2025Natur.642..905L}, 14 Her c: \citealt{2025ApJ...988L..18B}). 

\par Among the coldest known directly imaged objects, GJ 504 b is a window into the atmospheres of planetary-mass companions at $T_\mathrm{eff}$ $\sim$ 500 K, much cooler compared to the majority of directly imaged planets ($T_\mathrm{eff}$ $\sim$1000 K). This allows us to probe atmospheric chemistry at effective temperatures closer to Jupiter ($\sim$130 K), and the previously mentioned cold planets, eps Ind b ($\sim$275 K) and 14 Her c (275--300 K). GJ 504 b orbits the G0V primary GJ 504 A (e.g., \citealt{2005ApJS..159..141V}) at a separation of $\sim43\,au$ \citep{kuzuhara2013} and was the first directly imaged planetary-mass companion discovered around a Sun-like star. The low effective temperature also leads to GJ 504 b being too faint to observe using most ground-based spectrographs, even though the system is in the solar neighborhood ($d\,\sim17.6$ pc; \citealt{gaiaedr3}). Therefore, the companion only has photometric measurements \citep{kuzuhara2013,janson2013,skemer2016,bonnefoy2018}. Despite a lack of spectroscopic data, several interesting aspects of this companion have been identified. With $J$ $\--$ $H$ = -0.23 mag \citep{kuzuhara2013}, it is bluer in the near-infrared compared to most known directly imaged exoplanets, while being redder than other brown dwarfs at similar temperatures and luminosities. Thus, it occupies a unique position in the color-magnitude diagram \citep{bonnefoy2018}. \cite{janson2013} discovered methane absorption in the atmosphere of GJ 504 b using Subaru/HiCIAO \citep{2010SPIE.7735E..30S}, reaffirming its position as a T-type object.

\par The uncertainty around the age of the GJ 504 system has led to a debate over the nature of the companion as either a young planetary-mass object or an older brown dwarf. The lack of a dynamical mass for GJ 504 b implies that its luminosity is the only observable which can be compared to evolutionary models, with the degeneracy between mass and age at a specific luminosity leading to the two possibilities. \cite{kuzuhara2013} used gyrochronology and stellar activity indicators to arrive at an age of 160$^{+350}_{-60}$ Myr for GJ 504 A. Subsequent works have suggested accretion of a substellar object as the reason for its enhanced rotation and activity \citep{2015ApJ...806..163F, 2025A&A...694A.179P}, with revised age estimates of 1.8--3.5 Gyr using isochronal studies \citep{dorazi2017}. Further isochronal studies by \cite{bonnefoy2018} found two solutions for the age, with an estimate of 21 $\pm$ 2 Myr for a young system and 4.0 $\pm$ 1.8 Gyr for an old system.  These two ages lead to evolutionary mass estimates ranging from 1.3$^{+0.6}_{-0.3}\,M_\mathrm{Jup}$ (young) to 23$^{+10}_{-9}\,M_\mathrm{Jup}$ (old) for GJ 504 b. Among recent literature, \cite{2022ApJ...940...93D} and \cite{2023A&A...671A..10S} failed to break the age degeneracy, while work by \cite{2025A&A...694A.179P,2026arXiv260222979B} have found concrete evidence for an age 2.11 $\pm$ 0.46 Gyr for GJ 504 A by analyzing the star's rotation, X-ray luminosity, and magnetic activity. Irrespective of its age, the super-stellar metallicity of its companion from various atmospheric analysis \citep{skemer2016,bonnefoy2018,malin2025} have suggested that GJ 504 b may have formed like a planet.

\par In this work, we use medium-resolution ($R$ $\sim$ 2,700) 2.9--5.3 $\mu m$ $JWST$/NIRSpec data to perform a detailed spectroscopic analysis of this GJ 504 b, to measure its atmospheric properties including the effective temperature, surface gravity, metallicity, and elemental abundances. In Section \ref{sec:nirspecobs}, we outline the NIRSpec observations before proceeding to the data reduction and spectral extraction procedure in Section \ref{sec:datared}. We showcase detection of the companion using both a forward modeling analysis as well as the novel demonstration of PSF subtraction using Angular Differential Imaging (ADI) for the NIRSpec Integral Field Unit (IFU). We introduce our atmospheric retrieval framework in Section \ref{sec:atmosretrieval}, and  discuss our results in Section \ref{sec:gj504bresults}. We discuss the implications of our results to the GJ 504 b age/mass debate and its formation pathways in Section \ref{sec:discuss}, before presenting our conclusions in Section \ref{sec:conclusions}.

\par This paper is part of a series by the $JWST$ Telescope Scientist Team (JWST-TST). This collaboration uses Guaranteed Time Observations (GTO, PI: M. Mountain) for projects across three topic areas: Exoplanet and Debris Disk High-Contrast Science (lead: M. Perrin), Transiting Exoplanet Spectroscopy (lead: N. Lewis), and Local Group Proper Motion Science (lead: R. van der Marel). Our previous JWST-TST work in the High-Contrast area includes \cite{2024AJ....167...69R}, \cite{2024AJ....168...51K}, \cite{ruffio2024}, \cite{2024AJ....168..187H}, \cite{balmer2025}, and \cite{2025A&A...704A.181M}.

\section{Nirspec Observations}\label{sec:nirspecobs}

\begin{figure*}
    \centering
    \includegraphics[width=1.0\linewidth]{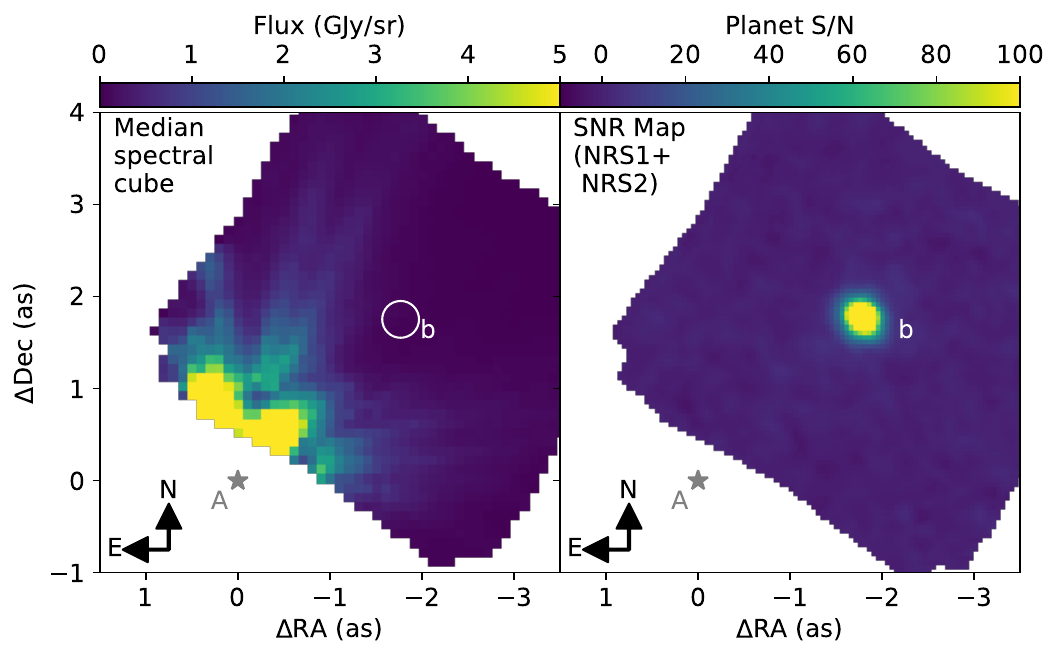}
    \caption{Detection map of companion GJ 504 b around the primary GJ 504 A using the medium resolution IFU mode of $JWST$/NIRSpec in the 2.9--5.3 $\mu m$ range. (\textit{Left}) Median spectral cube before starlight subtraction generated using stage 3 of the JWST science calibration pipeline \citep{bushouse2024}. (\textit{Right}) Signal-to-noise (S/N) detection maps for GJ 504 using the \texttt{BREADS} forward modeling framework introduced in \cite{ruffio2024}. GJ 504 b is detected with a S/N of 357.}
    \label{fig:gj504bdetmap}
\end{figure*}

The GJ 504 system was observed using the $JWST$ NIRSpec IFU \citep{jakobsen2022, boker2022} on 16 February 2024 (UT) as part of the GTO program 2778, with the detailed observations outlined in Table \ref{tab:gj504bobs}. The program involved two 151-minute observations of GJ 504 b using the G395H grating and the F290LP filter. The observations (Obs. 1 \& 2) were taken at different roll angles (Aperture PA $62.7^{\circ}$ and $52.7^{\circ}$ respectively) to enable speckle subtraction using Angular Differential Imaging (ADI) (e.g., \citealt{2006ApJ...641..556M}). The astrophysical scene for the observations is depicted in Figure \ref{fig:gj504bdetmap} (\textit{Left}). Owing to the wide separation of the companion and the primary, centering GJ 504 b in the IFU field-of-view led to the core of the PSF of the host star being left out of the field of view.

\begin{deluxetable*}{ccccccc}
    \centering
    \tabletypesize{\normalsize}
    \tablewidth{0pt}
    \caption{Observations of Cycle 2 GTO program 2778 (PI: Perrin) from 16 February, 2024 (UT). \label{tab:gj504bobs}}
    \tablehead{\colhead{Obs} & \colhead{Grating Filter} & \colhead{Groups} & \colhead{$T_\mathrm{int}$ per exp.} & \colhead{Dithers} & \colhead{Total $T_\mathrm{int}$} & \colhead{Description} }
    \startdata
    1 & G395H/F290LP & 32 & 8 min & 9 & 72 min & in field, aperture PA $=62.7^{\circ}$ \\
    2 & G395H/F290LP & 32 & 8 min & 9 & 72 min & in field, aperture PA $=52.7^{\circ}$
    \enddata
\end{deluxetable*}

The brightness of the primary also prevents us from using it for target acquisition. Instead, we rely on the absolute pointing accuracy of $JWST$ (0.1$^{\prime\prime}$) after guide star acquisition \citep[e.g.,][]{ruffio2024}. The two observations are taken as a non-interruptible sequence to ensure 1) consistency in the PSF and 2) the grating wheel remains at the G395H position throughout. The latter eliminates any potential systematics from minor differences in the position of the mechanism, potentially leading to minor differences in the PSF. We adopt a medium (0.5$^{\prime\prime}$) cycling dither pattern with nine positions for enhanced spatial sampling while also ensuring that the planet remains in the IFU FOV within the pointing accuracy of $JWST$.

The wavelength gap in the spectral coverage of the two NIRSpec detectors around 4$\mu m$ has different ranges depending on how the various IFU slices are projected onto the detector. This wavelength gap in the obtained science spectrum can be reduced through a series of offset exposures in which the target is observed across various IFU slices. Such exposures were not the goal of the program, which aimed to have the widely separated planet in the IFU FOV across all dithers while ensuring the core of the bright stellar PSF remained outside the field of view. This leads to a significant wavelength gap of $0.072\,\mu m$.

\section{Data Reduction and Spectral Extraction} \label{sec:datared}

\subsection{Flux calibration of detector images}
Typical NIRSpec IFU data reduction is performed using the $JWST$ data reduction pipeline \citep{bushouse2024} in multiple stages. Stage 1 processes the up-the-ramp readouts in the uncalibrated files (\textit{``*\_uncal.fits"}) to obtain the rate maps (\textit{``*\_rate.fits"}) in units of DN/s. Stage 2 of the pipeline then uses the rate maps to produce the flux-calibrated 2D images (\textit{``*\_cal.fits"}). Stage 3 of the pipeline then produces spectral cubes (\textit{``*\_s3d.fits"}) containing data that is spatially and spectrally regularly sampled. However, we do not use all stages of the JWST calibration pipeline, instead using the stage 2 outputs (the flux-calibrated detector images) as the starting point for further analyses.

The uncalibrated detector images were generated using version `2023\_4a' of the $JWST$ Science Data Processing software (SDP). We download the data from the Barbara A. Mikulski Archive for Space Telescopes (MAST; \citealt{2018SPIE10704E..13M}). Subsequently, stages 1 and 2 of the $JWST$ science calibration pipeline version `1.12.5' were used to obtain the flux-calibrated detector images. The data reduction used version `11.17.16' of the Calibration Reference Data System (CRDS) selection software with CRDS context version `jwst\_1234.pmap' \citep{2016A&C....16...41G}.

Systematic vertical stripes are present in the rate maps (output of Stage 1) due to 1/f noise. We address this issue through an intermediate step between Stages 1 and 2 of the official pipeline. The approach involves the use of a column-wise spline to fit the background column-by-column \citep{ruffio2024}. The instrument class \texttt{jwstnirpsec\_cal} in the open source repository \texttt{breads}\footnote{https://github.com/jruffio/breads} (Broad Repository for Exoplanet Analysis, Discovery, and Spectroscopy; \citealt{agrawal2024,ruffio2024}) is used to process the flux-calibrated detector images before PSF subtraction and spectral extraction. The steps are outlined in detail in \cite{ruffio2024}, and we describe them briefly here. 

The first step involves additional bad pixel identification and masking (apart from those previously identified using the ``DQ" extension in the ``cal" files) using row-by-row sigma clipping of the error maps.  The pixel sky coordinates are then converted from absolute Right Ascension (RA) and Declination (Dec) coordinates (RA$_{\mathrm{pix}}$, Dec$_{\mathrm{pix}}$) to offsets in arcseconds ($\Delta$RA$_{\mathrm{pix}}$, $\Delta$Dec$_{\mathrm{pix}}$) relative to the host star coordinates (RA$_{\mathrm{star}}$, Dec$_{\mathrm{star}}$). The stellar centroid in the data is offset by up to 0.16$^{\prime\prime}$ in RA and 0.09$^{\prime\prime}$ in declination compared to the WCS headers. To account for this, we fit for the centroid of a \texttt{STPSF} \citep{2012SPIE.8442E..3DP,2014SPIE.9143E..3XP} model for the combined set of dithers in each roll. The additional calibration offset in the stellar centroid is taken as the median of the offset across all wavelengths in each detector. This approach is viable as the variation in the centroid position with wavelength is $<$0.02$^{\prime\prime}$, being worse only at the edges of the detectors. 


\subsection{Computing the high-pass filtered planet spectrum}
\label{sec:hpf}

The subtraction of the stellar PSF is greatly complicated by the spatial undersampling of NIRSpec, resulting in the extracted point source spectrum not being photon-noise-limited \citep{law2023}. We use a spline model in the detector frame \citep{Ruffio2026} to fit and subtract the starlight at all locations in the detector images, making it possible to obtain a planetary spectrum with reduced systematics. However, this comes at the cost of losing the planet's continuum information.  As this process does not account for the planetary signal, we effectively obtain a high-pass filtered planetary spectrum that still contains the high-resolution spectral features.

The rows of the detector are not co-linear with the spatial direction, implying that the starlight continuum along a row has variations due to wavelength as well as the spatial direction. The curvature of the spectral traces of NIRSpec on the detector leads to larger variations in starlight intensity and stronger starlight residuals towards the left side of NRS1 and the right side of NRS2. \cite{ruffio2024} first identified these residuals and attributed them to the curvature of the trace and the limited number of spline nodes. Hence to minimize the residual starlight, we opt for a spline model with 40 nodes per detector row, but with different spacing between the nodes \citep{Ruffio2026}. The wavelength range of the detector is divided into four equal sections. For NRS1, the node spacings are 0.02, 0.03, 0.04, and 0.06 $\mu$m for sections going from left to right. It is the reverse for NRS2. This increase in node density towards the left edge of NRS1 and the right edge of NRS2 enables us to be aggressive in removing the starlight where required without affecting the planet signal in other parts of the detectors. The detection maps after starlight subtraction are shown in Figure \ref{fig:gj504bdetmap} (\textit{Right}).

After starlight subtraction, the spline-subtracted detector images are interpolated onto a regular wavelength grid according to the procedure in \cite{ruffio2024}. This wavelength interpolation is done for easier PSF fitting. The minimum wavelength, maximum wavelength and bin-size for this regularly sampled grid is 2.8595 $\mu m$, 4.1013 $\mu m$ and 6.8 $\times$ 10$^{-4}$ $\mu m$ for NRS1 and 4.0813 $\mu m$, 5.2787 $\mu m$, and 6.7 $\times$ 10$^{-4}$ $\mu m$ for NRS2. The planet spectrum is then extracted by fitting a \texttt{STPSF} model to the combined point cloud of all nine dithers for both roll angles at all wavelengths and a regular spatial coordinate grid with 0.01$^{\prime\prime}$ spacing. The absolute flux calibration of the \texttt{STPSF} model is done similarly to \cite{ruffio2024}. This process gives us two ``spectral cubes", one for each observatory roll.  We thus obtain two planetary high-pass filtered spectra, one for each observatory roll. The cubes are combined using a weighted mean and a 5$\sigma$ mask is applied to reject spatial locations with inconsistent flux estimates between the two rolls.  The combined spectral cube is then used to extract the planet fluxes at its projected separation, using relative astrometry from \cite{whereistheplanet}. The latter utilize orbital parameters for GJ 504 b from \cite{2020AJ....159...63B} for their calculations.

The flux error is estimated as the standard deviation of the flux across all spatial locations in an annulus 0.2$^{\prime\prime}$ wide at the same separation from the primary as the companion. A disk of radius 0.3$^{\prime\prime}$ centered on the companion location is masked to avoid biases in the flux errors. We also compute the covariance matrix for the spline-filtered spectrum, with the procedure highlighted in Appendix \ref{appendix:hpfacf}.

\begin{figure*}
    \centering
    \includegraphics[width=1.0\linewidth]{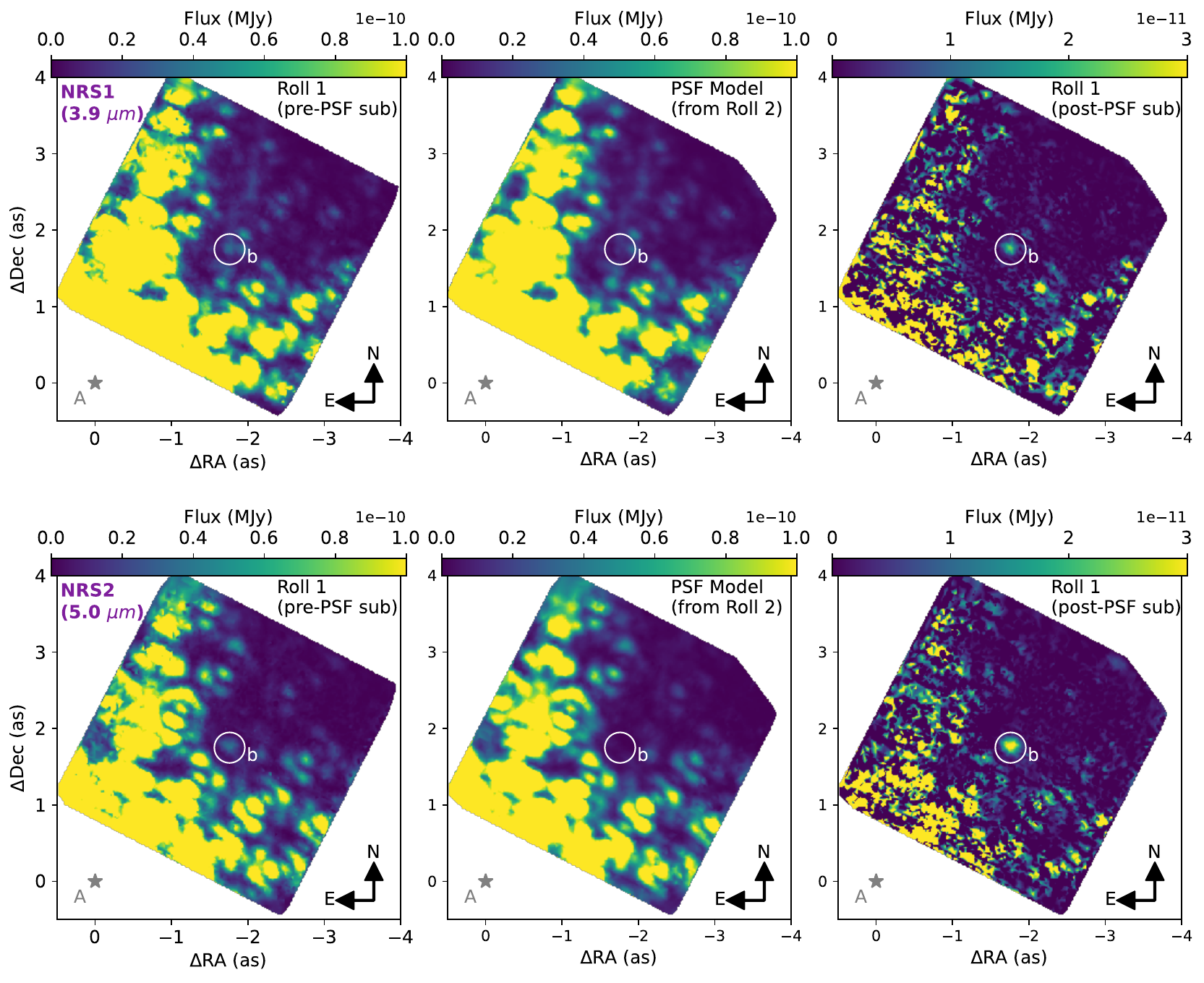}
    \caption{Different stages of PSF subtraction using ADI, shown for NRS1 (\textit{Top panel}) and NRS2 (\textit{Bottom panel}). Each image involves linear interpolation of the 2D point cloud and is shown only for visualization purposes. In each panel, the left-most image shows the flux-calibrated detector image from roll 1 before PSF subtraction. The middle image shows the reference PSF model generated from roll 2 after masking the companion. Right-most image depicts the detector images after PSF-subtraction, with the companion clearly detected at its marked position. The PSF centroid of the host star (GJ 504 A) is also shown.}
    \label{fig:gj504badi}
\end{figure*}

\begin{figure*}
    \centering
    \includegraphics[width=0.9\linewidth]{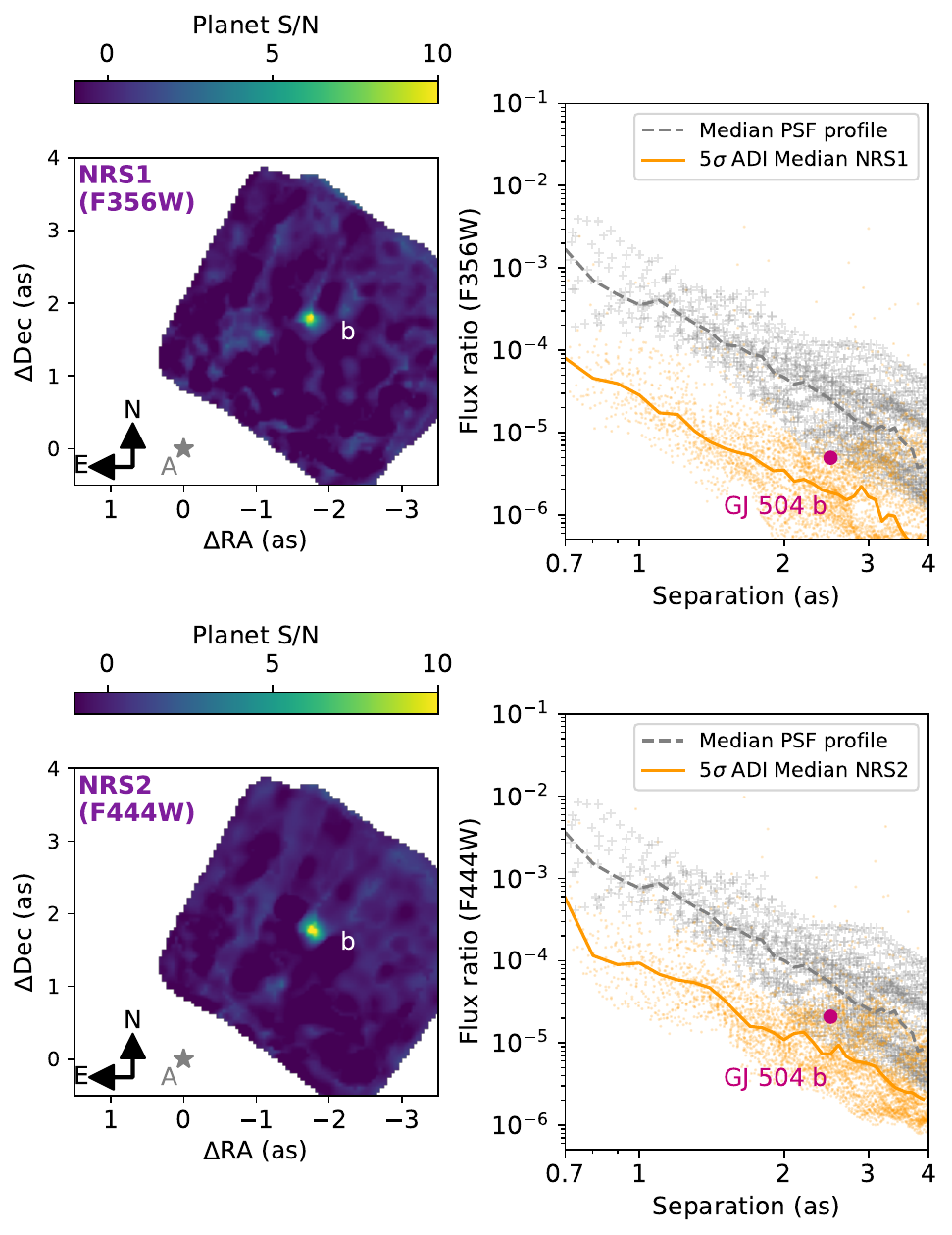}
    \caption{SNR maps and contrast curves for PSF subtraction using ADI, shown for NRS1 (\textit{Top panel}) and NRS2 (\textit{Bottom panel}). In each panel, the left-hand figure shows the SNR map, with GJ 504 b detected at an SNR$>10$ after renormalization of the NIRSpec fluxes to the F356W (NRS1) and F444W (NRS2) filters respectively. The right-hand figure shows the 5-sigma contrast curve. The gray dashed line indicates the median stellar PSF profile, with the 5$\sigma$ median contrast curve marked as the solid orange line. The gray scatter points depict the variability in the stellar PSF from NIRSpec at different position angles, leading to the scatter in the contrast limits (orange scatter). The orange scatter does \textit{not} depict the uncertainty in the 5$\sigma$ contrast limits. Contrast ratio of GJ 504 b is marked as the magenta point.}
    \label{fig:gj504badisnr}
\end{figure*}

\begin{figure*}
    \centering
    \includegraphics[width=0.9\linewidth]{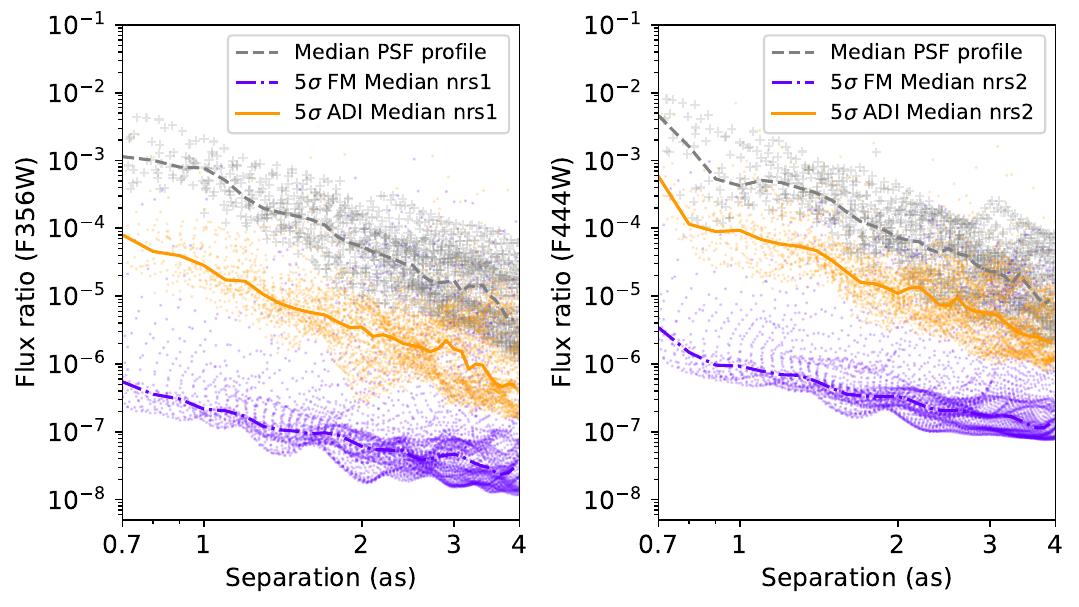}
    \caption{Sensitivity limits for ADI (orange solid line) and FM (violet dot-dashed line), expressed in the F356W filter for NRS1 and the F444W filter for NRS2. The gray dashed line indicates the median stellar PSF profile, with the gray scatter points showing the variability in the stellar PSF from NIRSpec at different position angles. Similar variability in the contrast limits for ADI and FM are shown as the orange and violet scatter points respectively. The contrast performance of ADI is two orders of magnitude worse than FM on average.}
    \label{fig:adivfm}
\end{figure*}

\subsection{PSF subtraction using Angular Differential Imaging (ADI)}

\subsubsection{Context}
While the spline modeling approach allows for a reasonably accurate subtraction of the starlight with limited systematics, we lose information regarding the planet continuum. The absolute continuum level of the planet spectrum is helpful for characterization of its atmosphere and also enables the identification of broad spectral features (e.g., CO$_2$ at 4.2--4.4 $\mu m$ in Fig. \ref{fig:gj504bmolecules}), which might be lost during the spline filtering process. However, obtaining the absolute flux level of the planet requires PSF subtraction to remove the starlight. This can be done either using a simulated model PSF (\texttt{STPSF}) or an empirical PSF. The former has no noise and minimal interpolation errors, but is subject to model imperfections. Meanwhile, an empirical PSF can accurately capture the systematics from the observation data, at the cost of limited S/N and greater interpolation errors. \cite{ruffio2024} demonstrated PSF subtraction using Reference Differential Imaging (RDI) to recover the continuum for the sub-stellar companion HD 19467 B. 
Angular Differential Imaging involves observations of the same science target at different telescope roll angles. Each roll angle is used as the science and reference observation in turn, allowing us to utilize our science time solely on observations of the target of interest. Telescope rolls also require less time compared to slewing to a reference star. The current program has NIRSpec observations at two different roll angles (Table \ref{tab:gj504bobs}), allowing us to utilize Angular Differential Imaging (ADI) to subtract the speckles and obtain the continuum-included planet spectrum. Our strategy involves performing ADI on the point cloud in order to avoid the large systematics in the datacubes \cite{law2023}. The point cloud is the collection of pixels across all valid detector images, irregularly sampled in both the spatial and wavelength directions. The irregular spatial sampling results from the projection of the IFU onto the detector. However, some degree of spatial interpolation is still required as the point clouds for the two observations rolls do not overlap perfectly.

\subsubsection{Generation and fitting of a reference PSF}

We utilize the same framework used by \cite{ruffio2024} in their RDI-based approach, but using an ADI roll to obtain the reference PSF instead of a different star. The two observations of the astrophysical scene at different roll angles were taken with the same instrument configuration, dithering strategies and exposure times. However, the rotated field of view implies PSF subtraction is simpler in the detector frame as compared to the sky frame. Since the speckle pattern in the flux-calibrated images is determined by the projection of the sky onto the detector at each wavelength, the speckles should be functionally identical in the detector frame (assuming that the speckle intensity at any point in the sky depends only on distance from the core of the stellar PSF). While this approach does ignore temporal variations in the speckle pattern and intensity, the stability of the $JWST$ wavefront (e.g., \citealt{ruffio2024}) allows us to consider them as negligible.

To generate the reference PSF corresponding to each observation, each detector image from that observation is interpolated onto a regular wavelength grid and the point cloud from all nine dithers are combined. The combined point cloud for each observation is then rotated by an angle of 215$^{\circ}$ to transform from the sky frame to the detector (i.e., IFU) frame for the PSF subtraction. We perform the PSF subtraction in concentric annuli that are 0.2$^{\prime\prime}$ wide. These are further divided into sectors, such that each sector has an area of approximately 0.5 arcsec$^2$. The companion centroid is kept fixed to prevent its position from being biased by adjacent speckles. The fitting of the reference PSF is done for each sector at each wavelength to obtain the final PSF subtracted detector images (Fig. \ref{fig:gj504badi}). These images are still interpolated onto the regular wavelength grid. We perform this procedure using each observation as the science and the reference PSF in turn, providing two sets of PSF subtracted, wavelength-interpolated detector images. 


\subsubsection{Spectral extraction \label{subsubsec:adispecextract}}

Spectral extraction from the PSF subtracted images is along similar lines to spectral extraction the high-pass filtered spectra described in Section \ref{sec:hpf}. First, we convert the PSF subtracted images from the detector frame to the sky frame. Next, we define a regular grid in sky coordinates with a spatial sampling of 0.01$^{\prime\prime}$ for both RA and Dec directions. A \texttt{STPSF} model is then fit at each position and wavelength to obtain the spectral cube with best-fit fluxes and errors. In addition to spectral extraction at the planet position (using relative astrometry from \citealt{whereistheplanet}), we also extract the spectra in an annulus with radii between 0.5--0.6$^{\prime\prime}$ about the planet position to sample the speckles. As the PSF subtraction using ADI is imperfect, the speckles have a non-zero median, which is subtracted from the planet spectra. These speckles have a wavelength dependence, hence we also compute a covariance matrix for the ADI spectrum (Appendix \ref{appendix:adiacf}). The matrix explicitly accounts for the wavelength dependence of the speckles. This ensures that the residual structure in the speckles does not greatly affect the atmospheric analysis if performed on the ADI spectrum.

The flux errors for the ADI spectrum at each wavelength are computed as the standard deviation of the speckle field inside the above annulus. Since we use each observation roll as the science and reference in turn, we get two sets of ADI spectra and flux errors. Next, we calculate the weighted mean of the two sets of spectra using the inverse of the square of the flux error at each wavelength as the weight. This `weighted mean' ADI spectrum is shown in Figure \ref{fig:gj504bmolecules}. The flux in the ADI spectrum is below zero near the methane absorption feature (3.1--3.6\,$\mu m$) due to the PSF subtraction procedure. Methane absorption leads to the companion flux being extremely low in this region, with the companion being much fainter compared the speckles.  As the companion flux is intrinsically very low in this regime, it is easier for imperfect speckle over-subtraction to result in negative values. At other wavelengths where the companion is brighter, there can be a small amount of over-subtraction without the fluxes becoming negative. In the speckle dominated regime of ADI, accurately recovering the companion flux at these wavelengths would require further improvements in the ADI methods and interpolation algorithms for NIRSpec data beyond the current state of the art we present here.

\subsubsection{Contrast limits and Sensitivity}
We use the fluxes and flux errors from the ADI-subtracted detector images to obtain the SNR maps and 5-$\sigma$ contrast limits. In particular, we wish to compare the performance of ADI to the \texttt{BREADS} forward-modeling (FM) detection framework. \texttt{BREADS} FM for $JWST$/NIRSpec normalizes the companion fluxes and flux errors to a reference NIRCam filter: F356W for NRS1 and F444W for NRS2 \citep{Ruffio2026}, before computing the SNR detection maps. However, the detection maps are obtained by fitting fluxes from the entire F290LP range, not just those corresponding to the NIRCam filter wavelengths. Hence, we perform a similar approach to compute the ADI sensitivity limits. First, the ADI fluxes obtained in Section \ref{subsubsec:adispecextract} are normalized to equivalent fluxes in the F356W (NRS1) and F444W (NRS2) filters. The equivalent fluxes are then interpolated onto the wavelength grid corresponding to the original ADI spectrum using \texttt{scipy.interpolate.interp1d}, to obtain the 1D template. Next, we fit the 1D template at each pixel in the ADI-subtracted spectral cube using the matched filter approach (e.g., \citealt{2017ApJ...842...14R}), giving us the flux map and error map for the companion. On computing the ADI SNR maps (Figure \ref{fig:gj504badisnr}), we find that GJ 504 b is detected at a SNR~=~14.7 in NRS1 and SNR~=~28.8 in NRS2.


As discussed in \cite{ruffio2024} for RDI, this matched filter approach does not account for covariances, leading our SNR detection map to be speckle noise dominated. Hence, we rescale the error and the SNR map using the standard deviation of the SNR map in a broad annulus $0.5^{\prime\prime}$--$1.0^{\prime\prime}$ about the position of GJ 504 b. The inner limit is to avoid the companion detection SNR biasing our rescaling factor. An outer limit is chosen to avoid the edges of the field of view where speckle subtraction has larger residuals due to the offset between the two rolls.

The contrast curves for ADI and FM are compared in Figure \ref{fig:adivfm}. On average, the contrast limit of FM is two orders of magnitude better than ADI. This is due to the residual speckles dominating the error budget for ADI-based subtraction, something that was also noted by \cite{ruffio2024} in their RDI-based approach. This highlights how the poor spatial sampling of NIRSpec combined with a simple linear interpolation scheme for the point cloud is limiting the performance of classical PSF subtraction.
The forward modeling approach and/or PSF-subtraction using much improved simulated PSF are required to maximize the high-contrast capabilities of the NIRSpec IFU. 

Regardless, the recovery of spectral continuum of the planet using ADI subtraction implies classical PSF subtraction cannot be entirely ignored, with both FM and ADI-based strategies having their own merits. In particular, accurate recovery of the continuum can enable better estimates of the companion luminosities due to more accurate bulk parameters (e.g., radius). An analysis strategy that utilizes both approaches enables recovery of the continuum (and accurate atmospheric parameters), while allowing detection and measurement of trace molecules. The use of NIRCam photometry, along with the NIRSpec FM mode is also a viable alternative to our strategy, with the photometry enabling some constraints on the continuum, and NIRSpec FM allowing for abundance measurements even for trace species (e.g., \citealt{Ruffio2026}). 



\section{Spectral analysis} \label{sec:atmosretrieval}

\subsection{Atmospheric Retrieval Framework}
This section describes the atmospheric retrieval framework used to characterize the PMC and infer its atmospheric properties. The radiative transfer code \texttt{petitRADTRANS} \citep{molli2019, molli2020} is used to generate synthetic spectra for GJ 504 b. The line-by-line molecular opacities, down-sampled from their native $R=$ 10$^6$ to $R=$ 100,000, are used to generate synthetic models for the $R\sim$ 2,700 NIRSpec G395H data. For photometric data, we use the correlated-k opacity mode down-sampled to $R$ = 100. The framework closely follows previous work by \cite{Xuan2026} and \cite{Ruffio2026}. Due to large noise covariance for the ADI spectrum and to reduce computational resources required, we choose to perform retrievals only on the high-pass filtered spectrum, rather than both of them. 

\begin{figure*}
    \centering
    \includegraphics[width=1.0\linewidth]{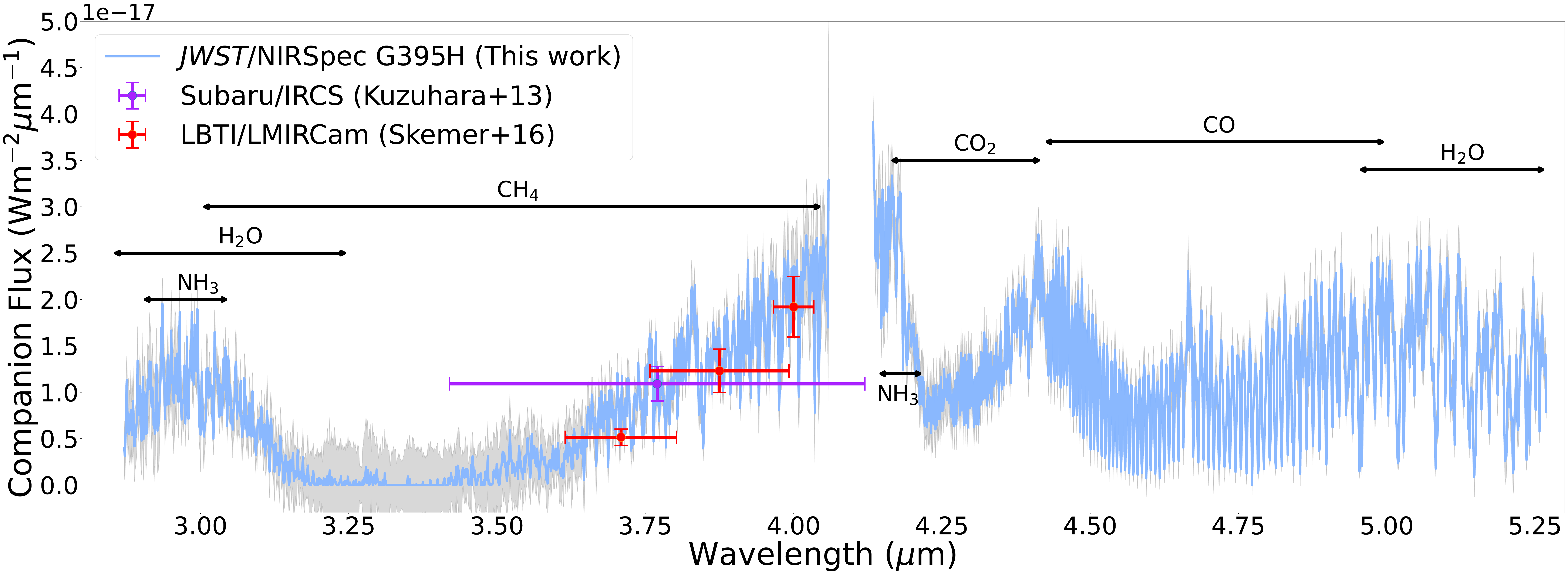}
    \caption{Post-ADI NIRSpec G395H spectrum of GJ 504 b (cyan) with the dominant molecular opacity sources $-$ H$_2$O, CO, CH$_4$, CO$_2$, and NH$_3$ marked. The gray shaded region indicates the 1$\sigma$ error on the extracted flux. Ground-based photometric points from \cite{kuzuhara2013} and \cite{skemer2016} are marked for reference.}
    \label{fig:gj504bmolecules}
\end{figure*}

\subsubsection{Opacity sources and Chemistry \label{subsubsec:opacitysource}}
Molecular opacities in our retrievals include $^{12}$C$^{16}$O (\texttt{HITEMP}; \citealt{hitemp}) and its two isotopologues, $^{13}$C$^{16}$O (\texttt{HITRAN}; \citealt{hitran} and $^{12}$C$^{18}$O (\texttt{HITRAN}), H$_2$O (\texttt{HITRAN}) and its isotopologue HDO (\texttt{HITEMP}), CH$_4$ \citep{hargreaves} and its isotopologue CH$_3$D (\texttt{HITRAN}), CO$_2$ (\texttt{HITEMP}), NH$_3$ (\texttt{HITRAN}), PH$_3$ \citep{saityph3}, H$_2$S (\texttt{HITRAN}), HCN \citep{harrishcn}, TiO \citep{molli2019}, VO \citep{molli2019}, FeH \citep{mollistfeh}, and C$_2$H$_2$ (\texttt{HITRAN}). We also include species Na \citep{allardna} and K (\citealt{VALD} for line-lists, \citealt{molli2019} for wings) with atomic line opacities. We use easyCHEM \citep{2024arXiv241021364L} to generate an equilibrium chemistry grid where abundances of all opacity species (except the isotopologues) are parametrized using the [C/H], [O/H], and [N/H] parameters (see details in \citealt{Xuan2026}). These parameters correspond to the atmospheric carbon, oxygen, and nitrogen abundances respectively. The abundances of other elements are scaled along the lines of the [C/H] parameter. The abundances of the isotopologues $^{13}$CO and C$^{18}$O are estimated using the CO abundance in the chemistry grid, and the $^{12}$CO/$^{13}$CO and C$^{16}$O/C$^{18}$O ratios respectively. Similarly, we use the H$_2$O abundance and the H$_2$O/HDO ratio, and CH$_4$ and the CH$_4$/CH$_3$D ratio respectively to interpolate for the HDO and CH$_3$D abundances respectively. All four isotopologue ratios are fitted parameters in our retrievals.  Lastly, we attempt a measurement of the sulfur abundance in the atmosphere of GJ 504 b. Under the assumption that the sulfur chemistry is mostly independent of the carbon chemistry, we fit for a H$_2$S scale factor, $f_{H_2S}$ (following \citealt{Xuan2026}), such that
\begin{equation}
    [S/H] = f_{H_2S} + [C/H]
\end{equation}

Our disequilibrium chemistry framework accounts for carbon and nitrogen quenching by fitting for a carbon quench pressure ($P_{\rm quench,C}$) and a nitrogen quench pressure ($P_{\rm quench,N}$) parameter. This prescription sets the abundances of CO, CH$_4$, H$_2$O, and HCN for $P$ $<$ $P_{\rm quench,C}$ to their values at $P=P_{\rm quench,C}$. We also account for ammonia quenching by defining a differential pressure ($P_\mathrm{diff}$), such that
\begin{equation}
    \log{P_{\rm quench,N}} = \log{P_{\rm quench,C}} + \log{P_\mathrm{diff}}    
\end{equation}
This treatment is on the basis of theoretical work suggesting ammonia quenching occurs at greater pressures (deeper in the atmosphere) compared to quenching of carbon-bearing species (e.g., \citealt{2022ApJ...938..107M}), and is identical to the approach outlined in \cite{Xuan2026}. The abundance of NH$_3$ for $P<P_{\rm quench,N}$ is set to its value at $P_{\rm quench,N}$. 

Atmospheric CO$_2$ abundances are not accurately described by the quenching timescale approximations, with recent work finding them to be constant with pressure for a significant portion of the atmosphere \citep{bieler2024, wogan2025}. Hence, CO$_2$ abundance is included as an additional free parameter in our retrievals by assuming its abundance to be vertically constant (invariant with pressure). Additionally, to avoid the overabundance of PH$_3$ predicted in models of cold substellar objects \citep{bieler2024, wogan2025}, we also fit for a constant-over-pressure PH$_3$ abundance. 

\subsubsection{Pressure-Temperature profile}
Our thermal structure extends from 10$^2$ to 10$^{-7}$ bars. From 10$^2$ to 10$^{-4}$ bars, we adopt the gradient pressure-temperature parametrization from \cite{zhang2023}, dividing the atmosphere into six layers equally spaced in logarithmic-pressure space. We fit for a temperature gradient $d\ln{T}/d\ln{P}$ for each layer, as well as the temperature at the bottom of the atmosphere ($T_{bottom}$). Normal priors are adopted for the gradients based on several self-consistent (i.e., in radiative-convective equilibrium) atmospheric model grids. We adopt the same priors for the temperature gradients as \cite{zhang2023}. Above these six layers, the P-T profile is assumed to be isothermal (from P$\,=10^{-4}$ to $10^{-7}$ bars).

\subsubsection{Forward Modeling}
Additional processing steps are applied to the \texttt{petitRADTRANS} synthetic spectra before comparisons to the science data. The first step involves applying a radial velocity (RV) shift to the model. Next, instrumental broadening is implemented by resampling the the model wavelengths such that the resultant wavelength grid aligns with the instrumental resolution. We follow the prescription by \cite{xuan2024c}, convolving the spectrum with a Gaussian kernel of standard deviation $\sigma$ which is defined as

\begin{equation}
\sigma = \frac{\lambda/R_\lambda}{2\sqrt{2\log{2}}}
\end{equation}

\noindent where $\lambda$ is wavelength. The term $R_\lambda$ approximates the wavelength-dependent instrumental resolution of the NIRSpec IFU and is obtained by interpolating the resolution against wavelength curve for the G395H grating from the official $JWST$ documentation\footnote{\href{https://jwst-docs.stsci.edu/jwst-near-infrared-spectrograph/nirspec-instrumentation/nirspec-dispersers-and-filters}{https://jwst-docs.stsci.edu/jwst-near-infrared-spectrograph}} onto the wavelength grid of the data using \texttt{numpy.interp}.  While fitting the high-pass filtered spectrum, there is also an additional step of applying the same spline-based high-pass filtering to the models.


After implementing these steps, we compare our model to the science spectra and calculate the residuals. The log likelihood is computed as follows

\begin{equation}
\ln{L} = -0.5 * \left[R^TC^{-1}R + n\ln{(2\pi)} + 0.5\ln{\det(C)}\right]
\end{equation}
\noindent
where $R$ is the residual array between the model and science spectra and $C$ is the covariance matrix. We also multiply $C$ by an additional error multiplicative term $N$ that we fit for as part of our retrievals.

\subsubsection{Photometric data}
In addition to the $R$ $\sim$ 2,700 NIRSpec spectra, we also fit previous Subaru \citep{kuzuhara2013, janson2013}, LBTI/LMIRcam \citep{skemer2016}, VLT/SPHERE \citep{bonnefoy2018}, and $JWST$/MIRI photometry \citep{malin2025}. Synthetic spectra are generated using the correlated-k opacities in \texttt{petitRADTRANS} down-sampled to $R$ = 100. The spectra are then radial velocity-shifted and fed to the SyntheticPhotometry function in the python package \texttt{species} \citep{species}, which generates synthetic photometry corresponding to each photometric point in our input data. The synthetic photometric points are compared to measured photometry to obtain the `photometric' log-likelihood. The final log-likelihood for each synthetic model is taken as the sum of the log-likelihoods from fitting the NIRSpec data and the photometric data.

\subsubsection{Model fitting}
The nested sampling package \texttt{dynesty} \citep{speagle2020} is used to sample the posteriors. We utilize 1000 live points, the `random walk' sampling method and set the stopping criterion to when the estimated contribution of the unsampled prior volume to the total evidence is $<$ 10\%. \texttt{dynesty} also provides the Bayesian evidence corresponding to each fit, allowing us to compare different models by calculating the Bayes factor $B$. $B$ is a quantitative measure of the likelihood of model $M_2$ compared to another model $M_1$. The details on the retrieved parameters and their priors are given in Table \ref{tab:param_prior}.

\begin{deluxetable*}{ll|ll}
\tablecaption{Retrieval parameters and their priors for GJ 504 b \label{tab:param_prior}}
\tabletypesize{\small}
\tablehead{
\colhead{Parameter} & \colhead{Prior} & \colhead{Parameter} & \colhead{Prior}
}
\startdata
Gravity ($\log{g}$)     & $\mathcal{U}(2.5,5.5)$              & $T_{\rm ref}$ [$P=10^{2}$] (K) & $\mathcal{U}(1000, 3000)$ \\
Radius ($R_\mathrm{Jup}$)      & $\mathcal{U}(0.6, 2.0)$             & $\left(d\ln{T}/d\ln{P}\right)_{1}$ & $\mathcal{N}(0.25, 0.025)$ \\
$[{\rm C/H}]$           & $\mathcal{U}(-0.5,1.5)$             & $\left(d\ln{T}/d\ln{P}\right)_{2}$ & $\mathcal{N}(0.25, 0.045)$ \\
$[{\rm O/H}]$           & $\mathcal{U}(-0.5,1.5)$             & $\left(d\ln{T}/d\ln{P}\right)_{3}$ & $\mathcal{N}(0.26, 0.05)$ \\
$[{\rm N/H}]$           & $\mathcal{U}(-0.5,1.5)$             & $\left(d\ln{T}/d\ln{P}\right)_{4}$ & $\mathcal{N}(0.20, 0.05)$ \\
$f_{H_2S}$              & $\mathcal{U}(-4.0,2.0)$                & $\left(d\ln{T}/d\ln{P}\right)_{5}$ & $\mathcal{N}(0.12, 0.045)$ \\
$\log{(^{12}\rm{CO}/^{13}\rm{CO})}$ & $\mathcal{U}(1, 4)$     & $\left(d\ln{T}/d\ln{P}\right)_{6}$ & $\mathcal{N}(0.07, 0.07)$ \\
$\log{(\rm{C}^{16}\rm{O}/\rm{C}^{18}\rm{O})}$ & $\mathcal{U}(1, 4)$  & $\log{(P_{\rm quench,C}/\rm bar)}$   & $\mathcal{U}(-6, 2)$ \\
$\log{(\rm{H}_2\rm{O}/\rm{HDO})}$\tablenotemark{b}   & $\mathcal{U}(1, 6)$            & $\log{(P_\mathrm{diff}/\rm bar)}$ & $\mathcal{U}(0, 2)$ \\
$\log{(\rm{CH}_4/\rm{CH}_3\rm{D})}$\tablenotemark{b}  & $\mathcal{U}(1, 6)$            & RV (kms$^{-1}$) & $\mathcal{U}(-50 , 50)$ \\
$\log{(\rm{CO}_2)}$ mass–mixing ratio & $\mathcal{U}(-12, 0)$        & Error factor $N$\tablenotemark{a} & $\mathcal{U}(1, 5)$ \\
$\log{(\rm{PH}_3)}$ mass–mixing ratio & $\mathcal{U}(-12, 0)$        & $\log{K_{zz}}$\tablenotemark{c}  & $\mathcal{U}(4, 13)$ \\
$\log{X_{0,\rm cloud}}$\tablenotemark{c} & $\mathcal{U}(-8, 0)$                     & $\log{f_{sed}}$\tablenotemark{c} & $\mathcal{U}(0, 10)$  \\
$\log{(P_{\rm base, cloud}/\rm bar)}$\tablenotemark{c} & $\mathcal{U}(-6, 2)$     & $\sigma_{g}$\tablenotemark{c} & $\mathcal{U}(1.05, 3)$      \\
\enddata
\tablecomments{$\mathcal{U}$ indicates that the priors are drawn from a uniform distribution, with the two values within parentheses indicating the end points of the distribution. $\mathcal{N}$ indicates that the priors are drawn from a normal distribution with the parameters within the parentheses indicating the mean and standard deviation of the distribution.
\tablenotetext{a}{$N$ is an error inflation term multiplied to the covariance matrices. We fit for it in all retrievals.}
\tablenotetext{b}{These parameters are fit only in clear atmosphere retrievals.}
\tablenotetext{c}{These parameters are fit only in cloudy atmosphere retrievals.}}
\end{deluxetable*}

\section{Results} \label{sec:gj504bresults}

\subsection{Molecular detections}
We report ``by-eye" detections of H$_2$O, CO, CO$_2$, CH$_4$, and NH$_3$ (Figure \ref{fig:gj504bmolecules}).
We perform free retrievals without the molecule of interest (leave-one-species-out retrievals) to validate the presence of CO$_2$, NH$_3$, H$_2$S, $^{13}$CO, and C$^{18}$O. This is done by comparing the Bayes factor from the leave-one-species-out or the `reduced' models to those of the `full' model. A negative Bayes factor for the leave-one-out reduced model compared to the full model indicates that the full model (i.e., model including the specific molecular opacity) is preferred. 

\par In addition to the Bayes factor, we perform a second validation test where we compute the cross-correlation function (CCF) between the residuals of the reduced model and the data, and a pure molecular template \citep[e.g.][]{2021Natur.595..370Z, xuan2024}. For example, in order to validate CO$_2$, we first calculate the residuals between the data and a reduced model with no CO$_2$. These residuals would contain CO$_2$ lines. Next, we compute the CCF using these residuals and a pure CO$_2$ molecular template, to obtain a CCF S/N on the detection of CO$_2$. The CCF analyses present an independent verification of molecular detections from the Bayes factor. The results of both of the above analyses are given in Table \ref{tab:moldetect}. 

Our detections include CO$_2$ at a $>80\sigma$ significance and NH$_3$ at a $\sim10\sigma$ significance. Among the wide-orbit companions, NH$_3$ has previously been detected in GJ 504 b using MIRI photometry \citep{malin2025}, the MIRI/LRS spectrum of the cold exoplanet WD 0806-661 b \citep{2025ApJ...982L..38V}, and a $4\sigma$ detection in HR 8799 b \citep{Xuan2026}. 
Other molecules detected include $^{13}$CO ($36\sigma$), C$^{18}$O ($12\sigma$), and H$_2$S ($6\sigma$). We report a non-detection of PH$_3$ with the positive Bayes factor from the leave-one-out analysis (Table \ref{tab:moldetect}). We only obtain an upper limit of $\log{\rm VMR} <-9.49$ (95\% posterior probability) from the baseline free retrieval (which includes all molecular opacities).

\begin{deluxetable*}{c|ccc}
    \centering
    \tablecaption{Molecular detections from leave-one-out retrievals and CCF analyses\label{tab:moldetect}}
    \tablewidth{0pt}
    \tablehead{\colhead{Molecule} & \colhead{\CellWithForceBreak{$\Delta\ln{B}$ \\ (compared to full model)\tablenotemark{1}}} & \colhead{\CellWithForceBreak{Detection significance \\ (from Bayes factor)}} & \colhead{\CellWithForceBreak{Detection significance \\ (from CCF)}}}
    \startdata
    CO$_2$ & -1051.4 & $45.9\sigma$ & $81.9\sigma$ \\
    $^{13}$CO & -472.7 & $30.7\sigma$ & $36.0\sigma$ \\
    C$^{18}$O & -68.4 & $11.7\sigma$ & $12.1\sigma$ \\
    NH$_3$ & -55.8 & $10.6\sigma$ & $9.5\sigma$  \\
    H$_2$S & -28.5 & $7.5\sigma$ & $6.0\sigma$  \\
    HDO & +0.7 & -- & \\
    CH$_3$D & +0.8 & --& \\
    PH$_3$ & +1.3 & -- & \\
    \enddata
    \tablenotetext{1}{Negative $\Delta\ln{B}$ indicates the full model is preferred compared to the reduced model (i.e., model without molecule of interest). Positive $\Delta\ln{B}$ indicates the reduced model is preferred over the full model.}
\end{deluxetable*}

\subsection{Clouds}
\begin{figure*}
    \centering
    \includegraphics[trim={0 0cm 0cm 0cm},clip,width=0.55\linewidth]{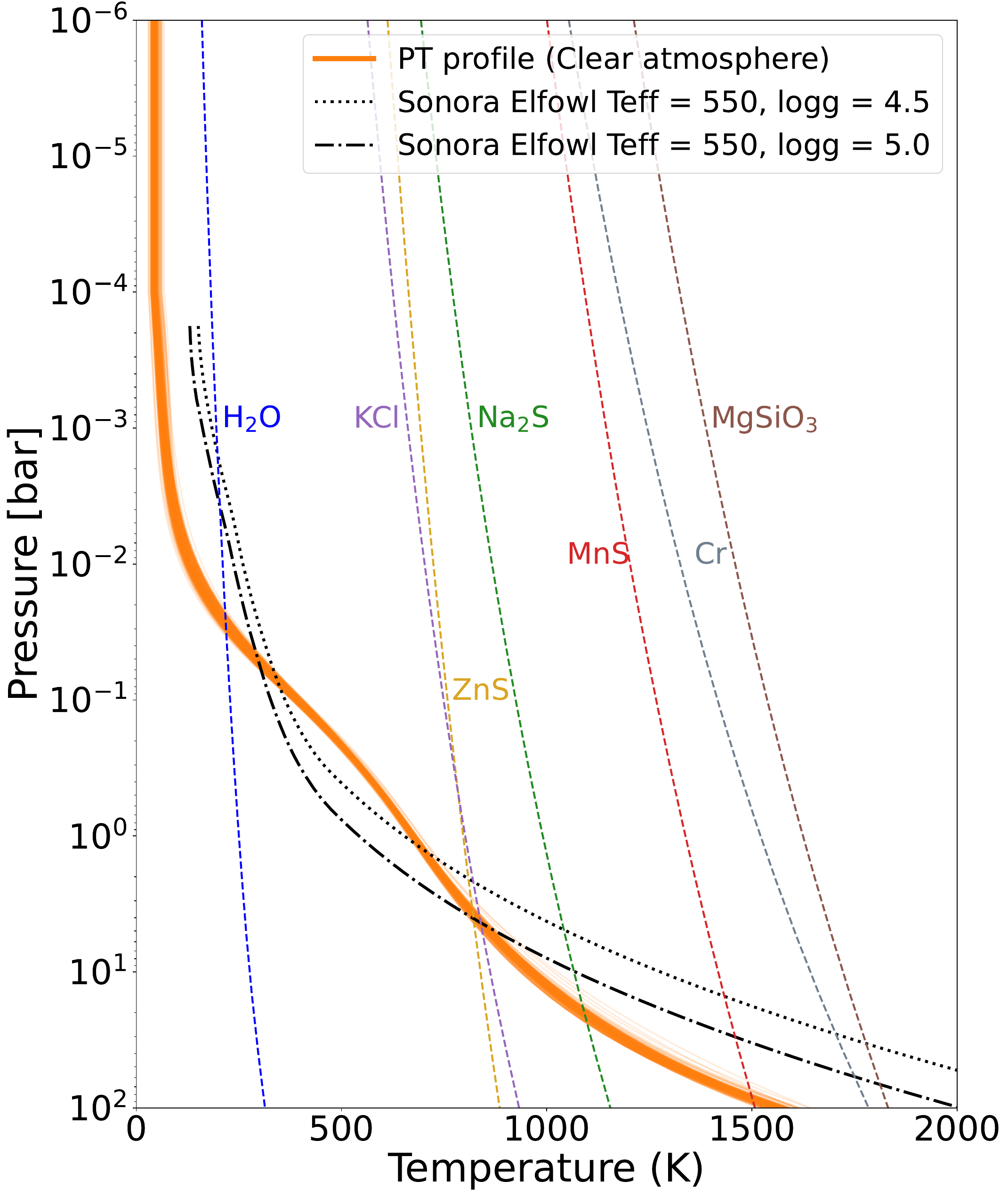}
    \caption{Pressure-temperature profile from a clear retrieval (no clouds) depicted using 200 randomly drawn P-T curves (orange) from the posterior chains. The P-T profile for a pre-computed radiative-convective equilibrium (RCE) model from the Sonora Elfowl grid (black dotted/dot-dashed lines; \citealt{2024ApJ...963...73M}) is shown for reference. The retrieved profile deviates from RCE around 0.1$-$1 bar indicating missing opacity sources in that region. The cloud condensation curves for different condensates are also indicated. KCl and ZnS clouds coincide best with the base of the isothermal region.}
    \label{fig:gj504bclearpt}
\end{figure*}

Conventional models of T-dwarf atmospheres assume a lack of clouds (e.g., \citealt{2008ApJ...689.1327S}) due to their bluer J-K colors compared to late-L dwarfs (e.g., \citealt{2006ApJS..166..585B, 2006ApJ...647.1393L, 2007ApJ...659..655B, 2015ApJ...805...56D}). However, \cite{2012ApJ...756..172M} suggested that dispersal of Fe and silicate clouds at the L/T transition could lead to species such as Na$_2$S, Cr, MnS, ZnS, and KCl becoming the dominant cloud condensates in the atmospheres of T-dwarfs. The initial sets of retrievals for GJ 504 b assumed a clear atmosphere (no clouds). However, these models give a retrieved thermal structure with a significant isothermal region around 0.1--1 bar (Figure \ref{fig:gj504bclearpt}), suggesting potentially missing opacity sources in our clear retrieval model. 

In order to test this hypothesis, we perform retrievals with an enstatite (MgSiO$_3$) cloud deck. The cloud properties are parameterized using the \texttt{EddySed} model proposed by \cite{eddysed}, as incorporated into \texttt{petitRADTRANS}. In this scheme, the cloud particle size is determined using the eddy diffusion coefficient ($K_{zz}$) which represents the strength of the vertical mixing and the cloud sedimentation parameter ($f_{sed}$). $f_{sed}$ is the ratio of the mass-weighted settling velocity of cloud particles and the local convective velocity, and effectively sets a scale-height for the cloud deck. The particle size distribution is assumed to be log-normal, with the width set using $\sigma_g$. Conventional EddySed implementations involve setting the cloud base pressure at the intersection of the P-T profile and the cloud condensation curves, which has been shown to be insufficient in modeling clouds \citep{2025arXiv250718691M}. Hence, we instead adopt a more flexible model where the cloud base pressure ($P_{base}$) is a fitted parameter for each cloud species. 

\begin{figure*}
    \centering
    \includegraphics[trim={0 0cm 0cm 0cm},clip,width=0.55\linewidth]{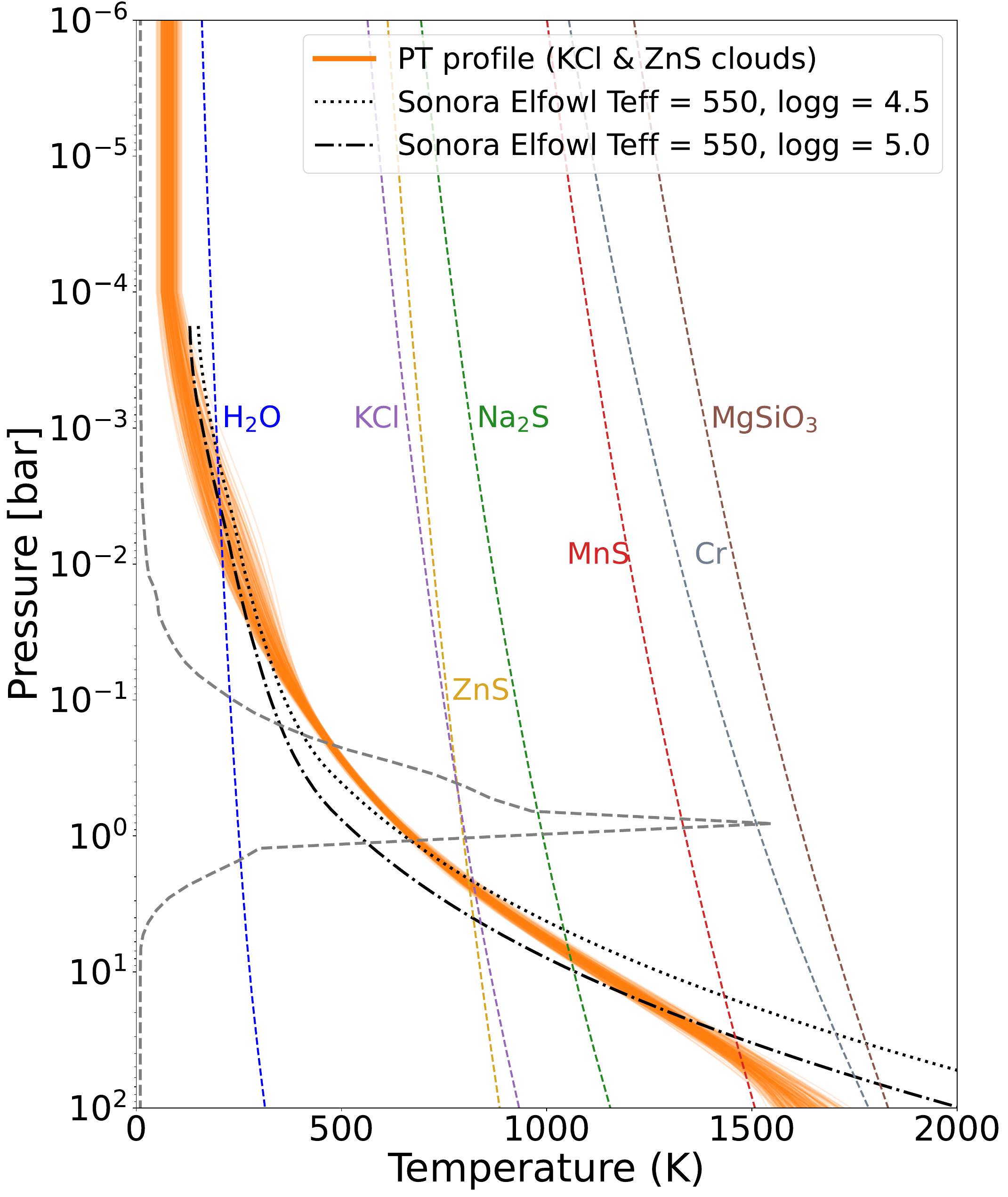}
    \caption{Same as Figure \ref{fig:gj504bclearpt} but for retrievals with KCl \& ZnS clouds. The additional opacity brings the P-T profile into better agreement with RCE models. We also indicate the cloud condensation curves for different cloud species. The gray dashed line indicates the wavelength-weighted contribution to the emergent flux at different pressures and highlights the preference for cloud base pressures $\sim1$ bar.}
    \label{fig:gj504bkclznspt}
\end{figure*}

Incorporation of enstatite clouds is successful in bringing the retrieved thermal profile in line with self-consistent models in radiative-convective equilibrium (RCE) by providing the necessary additional opacity needed to get rid of the isothermal region. However, at the temperatures of GJ 504 b, 1) silicate-based clouds should be below the observed photosphere, and 2) salt-based clouds should be the dominant condensate species. Hence, we run retrievals with a variety of different cloud species, including Na$_2$S, MnS, and KCl + ZnS. KCl \& ZnS clouds are the most natural choice of GJ 504 b, as the cloud condensation curves for those species intersect the clear atmosphere P-T profile at the base of the isothermal region (Fig. \ref{fig:gj504bclearpt}). Similar to enstatite, the retrieved thermal structure with these clouds also does not have an isothermal region (Figure \ref{fig:gj504bkclznspt}). 

\begin{figure*}
    \centering
    \includegraphics[trim={0 0cm 0cm 0cm},clip,width=1.0\linewidth]{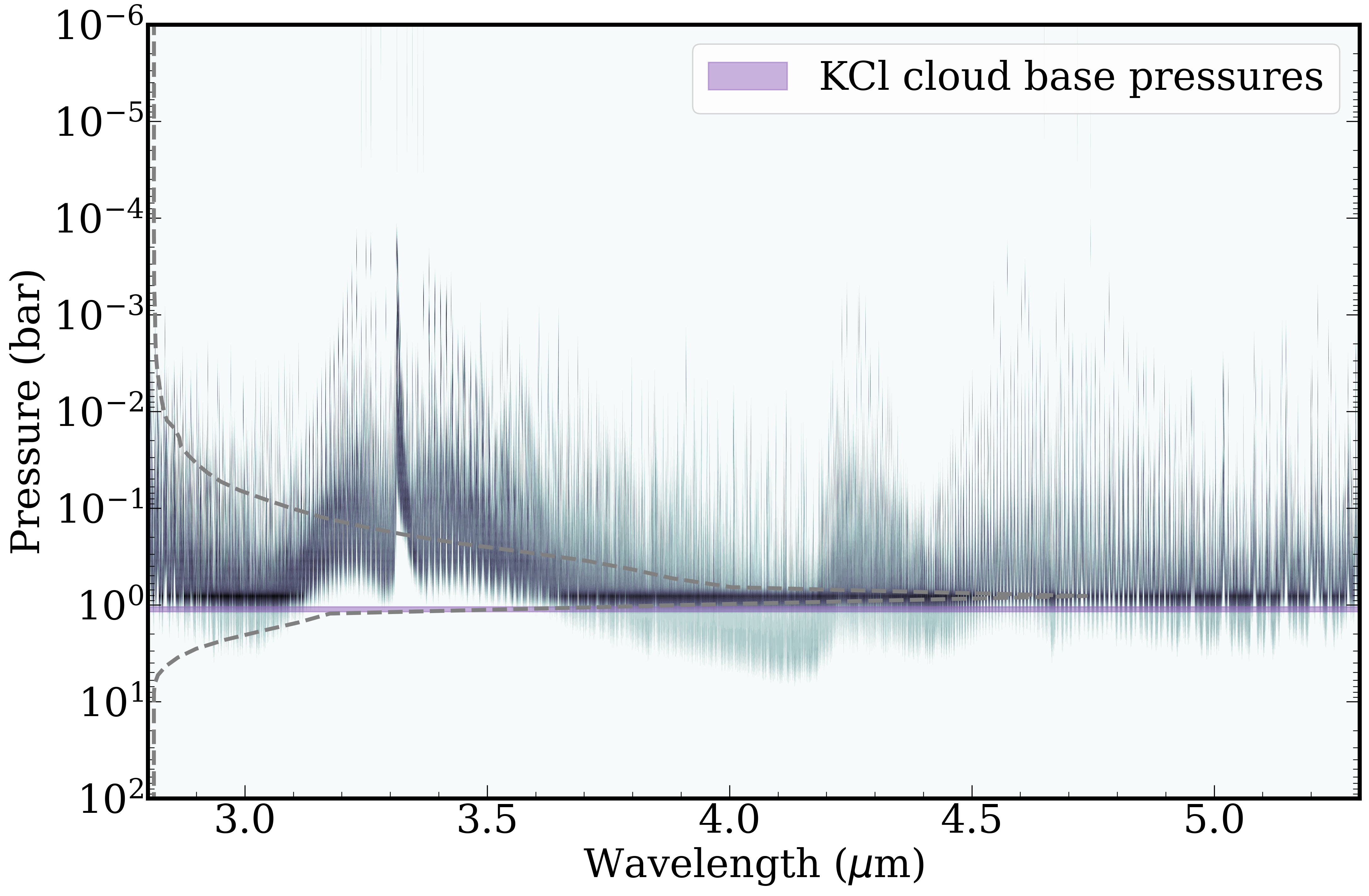}
    \caption{Emission contribution function for a cloudy retrieval with KCl \& ZnS clouds. The gray dashed line indicates the wavelength-weighted contribution function. The KCl cloud deck at $\sim1$ bar block the flux emerging from deeper in the atmosphere from reaching the surface, reducing the contribution of molecular opacities to the overall flux. The reduced contribution is also witnessed in the sharp decrease in the retrieved abundances between the clear and cloudy retrievals (Table \ref{tab:gj504bmanymodels}). The cloud base pressures and abundances of the ZnS cloud deck are poorly constrained from our retrievals (Figure \ref{fig:cloudyposteriors}), with the wavelength-weighted contribution function indicating it does not significantly impact the emergent flux. Hence, we do not mark it on this figure.}
    \label{fig:gj504bkclznsemission}
\end{figure*}

Salt clouds do not have spectral features at wavelengths 1--16$\,\mu m$ (e.g., \citealt{wakeford2015,nasedkin2024}). While the MnS cloud model is the most preferred model from our suite of retrievals (Table \ref{tab:gj504bmanymodels}), it is unclear why MnS clouds are preferred over KCl \& ZnS clouds. The cloud condensation curves indicate MnS clouds should form at pressures $\sim100$ bars, which is significantly below observable photosphere (Figure \ref{fig:gj504bkclznsemission}). Since the KCl + ZnS clouds are the more natural choice for the thermal structure of GJ 504 b, we adopt the latter cloud model as the fiducial model for further analysis. Note that this choice has minimal impact on the discussion in the rest of this work, as the retrieved bulk parameters (gravity, radius, luminosity), as well as the abundances are in excellent agreement for the MnS and the KCl + ZnS cloud models (Table \ref{tab:gj504bmanymodels}).

\begin{figure*}
    \centering
    \includegraphics[trim={0cm 0cm 0cm 0cm}, clip, width=1.0\linewidth]{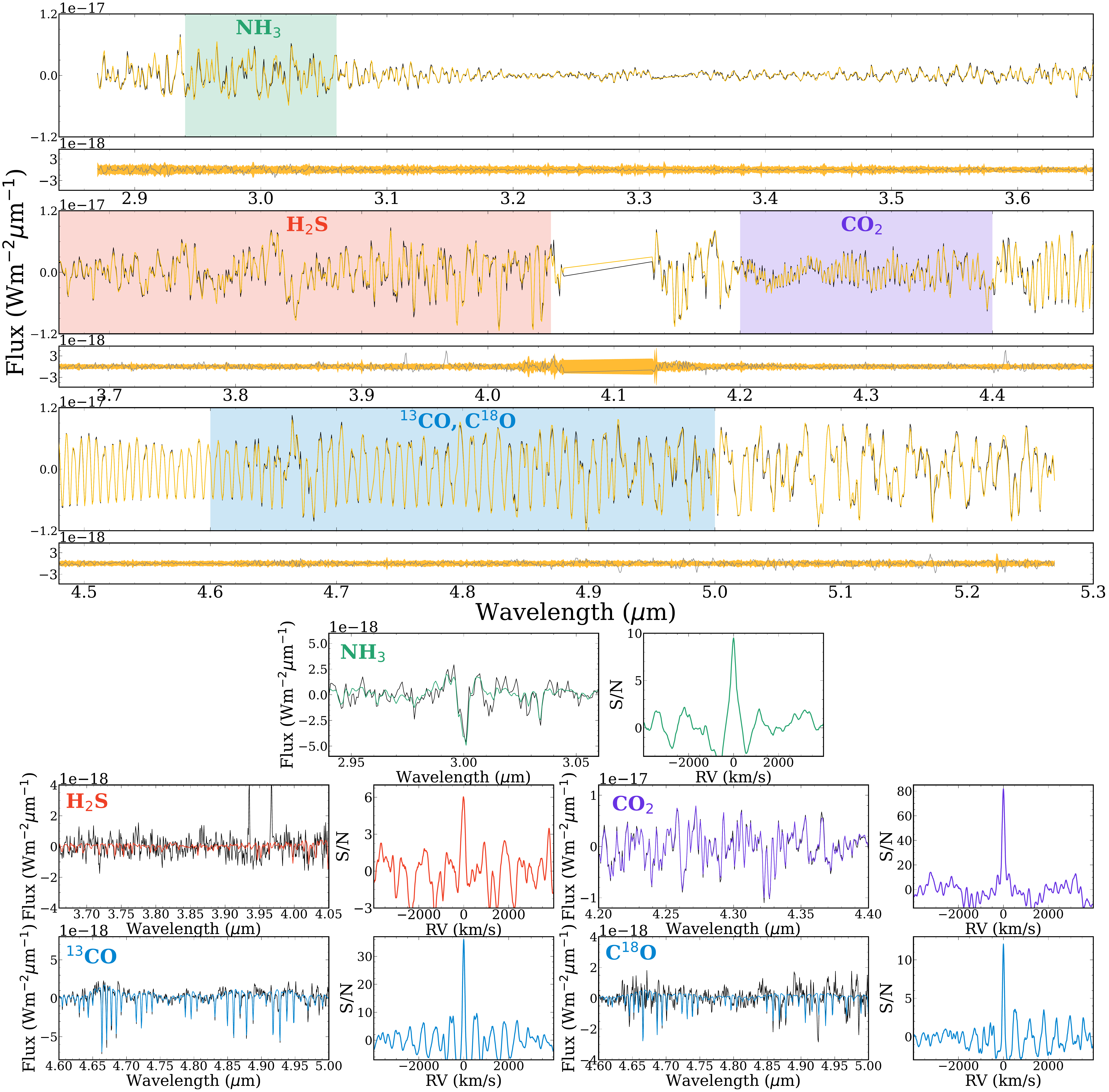}
    \caption{\textit{(Top)}: Best-fit spline-filtered cloudy model (KCl \& ZnS clouds) from \texttt{petitRADTRANS} (yellow) fit to the spline-filtered NIRSpec G395H spectrum (black) of GJ 504 b. The residuals between the model and data are shown in gray and the flux errors multiplied by the error inflation term are shown in gold. The shaded regions on the upper set of plots showcase the wavelength regions where trace species are detectable. \textit{(Bottom)}: For each shaded region (i.e., species) in the top figure, the plots on the left show the spectral signature of the specific molecule (in color) superposed on the residuals (in black) between the planet spectrum and the best-fit reduced model (model without species of interest), while the right plots show the detection signal-to-noise using the cross-correlation technique.}
    \label{fig:gj504bestfit}
\end{figure*}

\begin{figure*}
    \centering
    \includegraphics[trim={0cm 0cm 0cm 0cm}, clip, width=1.0\linewidth]{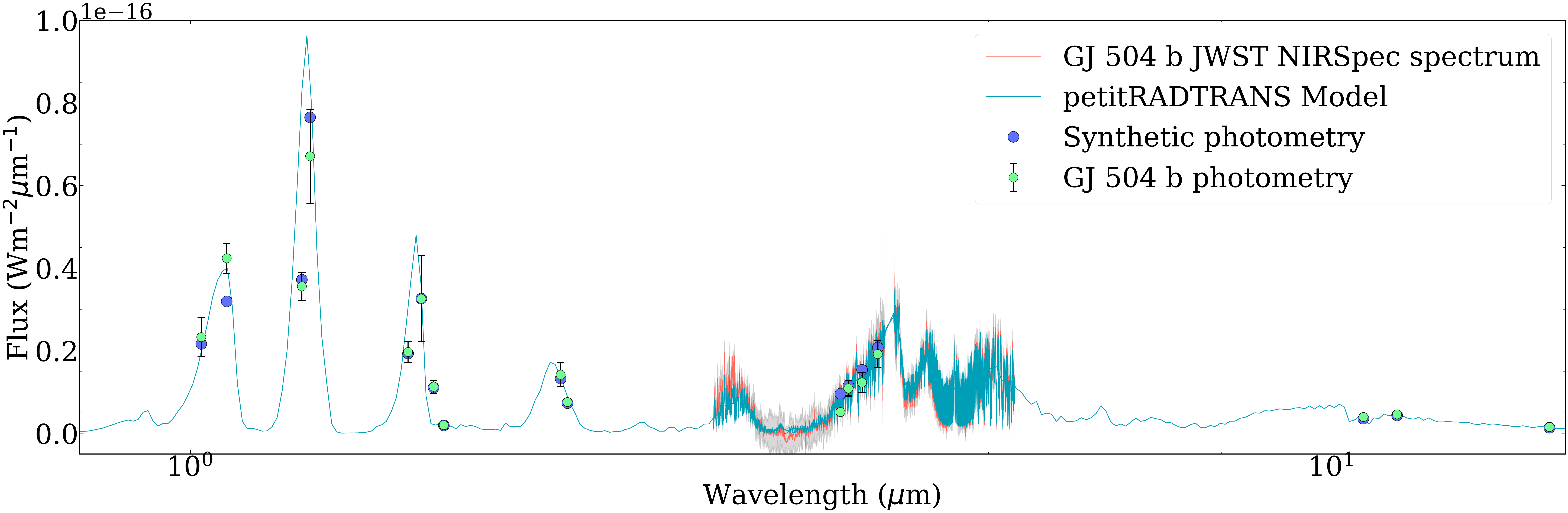}
    \caption{The same cloudy model (KCl \& ZnS clouds) as the spline-filtered spectra, but fit to the photometry (green points) and the ADI-subtracted NIRSpec spectrum of GJ 504 b (red). Gray indicate the 1$\sigma$ flux errors on the ADI spectrum. The blue points indicate the synthetic photometric points generated using the low-resolution model.}
    \label{fig:gj504photoandspecbestfit}
\end{figure*}

The fit of the KCl + ZnS cloudy model to the spline-filtered G395H spectrum of GJ 504 b is shown in Figure \ref{fig:gj504bestfit}. The exact same model is also overlaid on the spectrum extracted after ADI PSF subtraction, and photometric points, shown in Figure \ref{fig:gj504photoandspecbestfit}. It should be noted that this figure (i.e., Fig. \ref{fig:gj504photoandspecbestfit}) does \textit{not} represent the results of a retrieval fit to the spectrum extracted post ADI PSF subtraction.

\begin{figure*}
    \centering
    \includegraphics[trim={0 0cm 0cm 0cm},clip,width=1.0\linewidth]{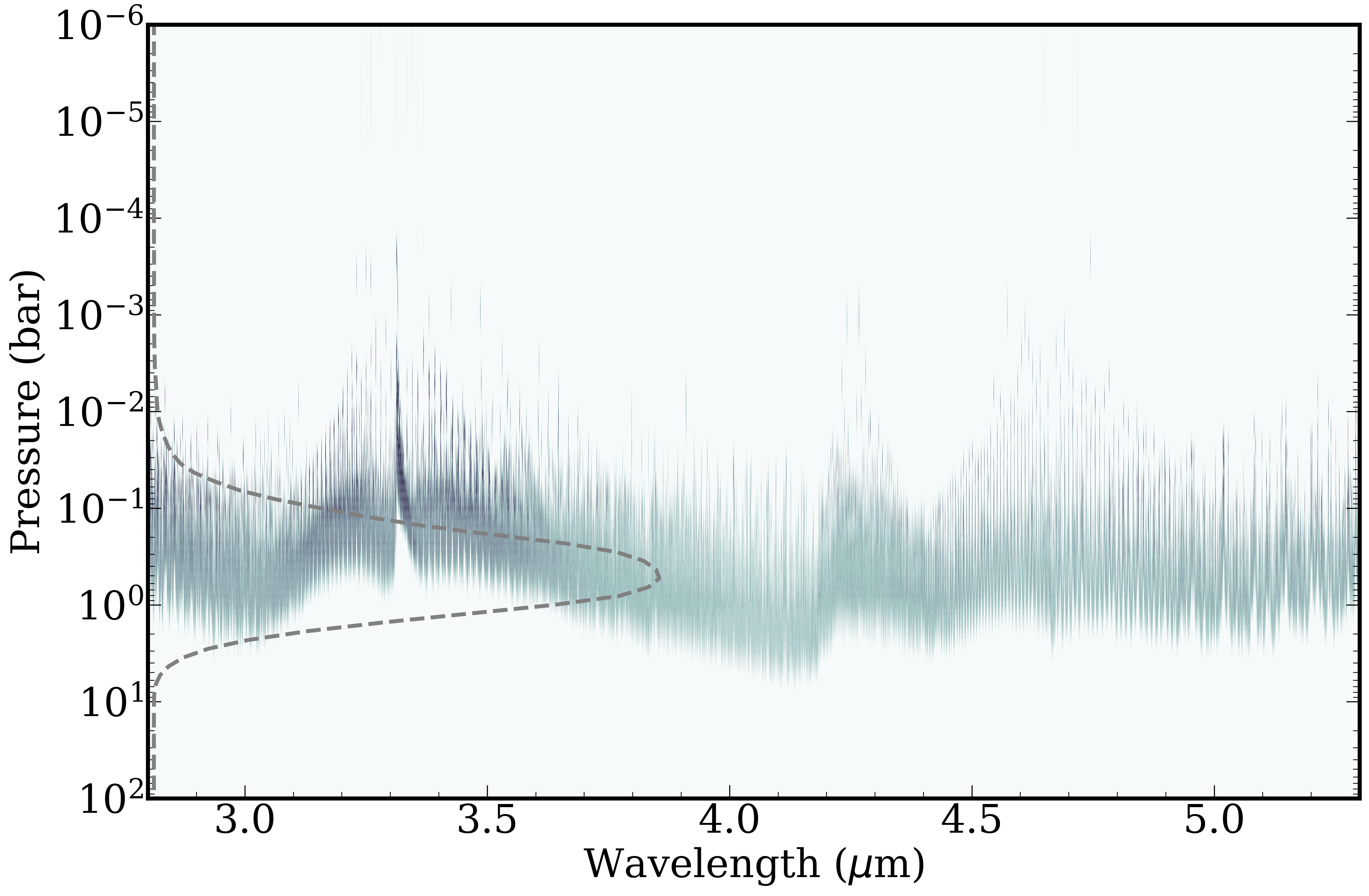}
    \caption{Emission contribution function for a clear atmosphere retrieval. These functions show the contribution to the emitted flux as a function of wavelength at different pressures in the atmosphere, with darker colors indicating greater flux contribution at those pressures. The gray dashed line indicates the wavelength-weighted contribution function. In clear atmospheres, there is increased flux from deeper layers of the atmosphere.}
    \label{fig:gj504bclearemission}
\end{figure*}

To understand the impact of clouds on the high-pass filtered spectrum, one can look at the emission contribution functions for clear (Figure \ref{fig:gj504bclearemission}) and cloudy retrievals (Figure \ref{fig:gj504bkclznsemission}). Clear atmosphere retrievals show emergent flux from deeper in the atmosphere In the presence of the KCl cloud deck at $\sim1$ bar, the emergent flux from deeper in the atmosphere is cut-off. This reduces the contribution of molecular opacities to the overall flux, an effect also seen in the sharp decrease in the retrieved abundances between the clear and cloudy retrievals (Table \ref{tab:gj504bmanymodels}). Without clouds, the additional opacity preferred by the data is provided by species from deeper in the atmosphere (i.e., at higher temperatures). This preference for additional molecular opacities also leads to the isothermal region in the P-T profile from clear retrievals (Figure \ref{fig:gj504bclearpt}). In cloudy retrievals, the cloud deck provides the increased opacity, making the additional opacities redundant and giving a P-T profile more aligned with RCE (Figure \ref{fig:gj504bkclznspt}).



\begin{deluxetable*}{c|ccccccccccc}
    \centering
    \tabletypesize{\scriptsize}
    \tablecaption{Retrieved parameters from different cloud prescriptions \label{tab:gj504bmanymodels}}
    \tablewidth{0pt}
    \tablehead{\colhead{Model} & \colhead{$\log{g}$} & \colhead{\CellWithForceBreak{Radius \\ ($R_\mathrm{Jup}$)}} & \colhead{$\log{L_{bol}/L_{\odot}}$} & \colhead{[C/H]} & \colhead{C/O Ratio} & \colhead{[N/H]} & \colhead{[S/H]} & \colhead{\CellWithForceBreak{$P_{\rm quench,C}$ \\ (bars)}} & \colhead{$^{12}$CO/$^{13}$CO} & \colhead{C$^{16}$O/C$^{18}$O} & \colhead{$\Delta\ln{B}$}}
    \startdata
    \CellWithForceBreak{Clear \\ (No clouds)} & $5.40^{+0.07}_{-0.09}$ & $0.90\pm0.02$ & $-6.10\pm0.01$ & $1.33^{+0.08}_{-0.11}$ & $0.64\pm0.02$ & $0.88\pm0.16$ & $1.26^{+0.10}_{-0.12}$ & $37.4^{+7.8}_{-1.7}$ & $71\pm4$ & $518^{+68}_{-60}$ & 0\\ \hline
    \CellWithForceBreak{MgSiO$_3$ \\ clouds} & $4.84^{+0.14}_{-0.08}$ & $0.92^{+0.02}_{-0.01}$ & $-6.09\pm0.01$ & $0.65^{+0.13}_{-0.09}$ & $0.68\pm0.01$ & $-0.34^{+0.29}_{-0.12}$ & $0.51\pm0.14$ & $58.0^{+14.1}_{-22.1}$ & $76^{+5}_{-6}$ & $695^{+140}_{-116}$ & +69.9 \\
    \CellWithForceBreak{MnS \\ clouds} & $4.92^{+0.12}_{-0.10}$ & $0.91^{+0.02}_{-0.02}$ & $-6.10\pm0.01$ & $0.71^{+0.11}_{-0.09}$ & $0.64\pm0.02$ & $-0.12^{+0.25}_{-0.23}$ & $0.60^{+0.14}_{-0.12}$ & $23.2^{+8.3}_{-3.5}$ & $72\pm4$ & $538^{+72}_{-69}$ & +95.9 \\
    \CellWithForceBreak{Na$_2$S \\ clouds} & $5.06^{+0.30}_{-0.23}$ & $0.89\pm0.01$ & $-6.11\pm0.01$ & $0.95^{+0.34}_{-0.25}$ & $0.64\pm0.01$ & $0.62^{+0.42}_{-0.39}$ & $0.82^{+0.33}_{-0.24}$ & $7.1\pm0.3$ & $72\pm4$ & $447^{+64}_{-59}$ & +84.9 \\
    \CellWithForceBreak{KCl + ZnS \\ clouds} & $4.87^{+0.13}_{-0.12}$ & $0.92\pm0.02$ & $-6.09\pm0.01$ & $0.67^{+0.13}_{-0.12}$ & $0.64\pm0.02$ & $-0.13^{+0.27}_{-0.24}$ & $0.59^{+0.15}_{-0.14}$ & $16.8^{+1.1}_{-1.0}$ & $72^{+5}_{-4}$ & $497^{+71}_{-61}$ & +91.5
    \enddata
\end{deluxetable*}

\subsection{Disequilibrium chemistry}
\begin{figure}
    \centering
    \includegraphics[width=1.0\linewidth]{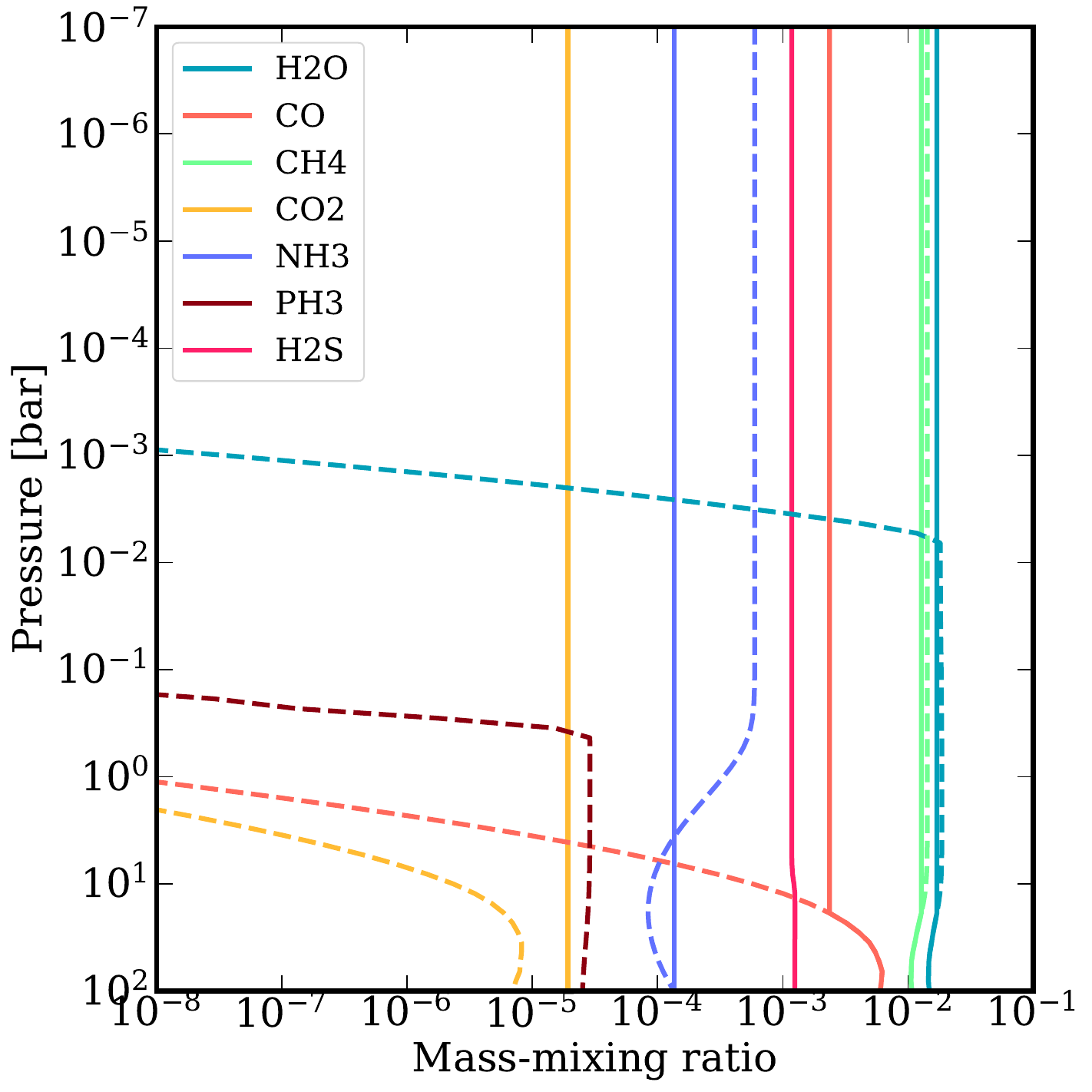}
    \caption{Retrieved mass-mixing ratios (MMR) with pressure for molecules in the atmosphere of GJ 504 b. The solid curves are under the assumptions of carbon and NH$_3$ disequilibrium chemistry and a constant-over-pressure CO$_2$ and PH$_3$. Dashed lines indicate the MMR profiles under chemical equilibrium, which would under-predict the CO and CO$_2$ abundance, and over-predict the CH$_4$, NH$_3$, and PH$_3$ abundance.}
    \label{fig:mmrprofile}
\end{figure}

We constrain the carbon quench pressure ($P_{\rm quench,C}$) to $16.8^{+1.1}_{-1.0}$ bars for GJ 504 b, which is significantly below the observable photosphere. The quench pressure is defined as the pressure at the point where the chemical reaction timescale ($\tau_{chem}$) equals the mixing timescale ($\tau_{mix}$). The retrieved quench pressure value indicates that abundances of H$_2$O, CO, and CH$_4$ are vertically constant over most of our adopted pressure range. The mixing timescale is used to defined the vertical diffusion coefficient ($K_{zz}$) via the relation $\tau_{mix}=L^2/K_{zz}$, where $L$ is the mixing length scale. We adopt the definition used by \cite{zahnle2014} and set $L$ = $H$, where $H$ is the pressure scale-height, thus getting $\tau_{mix}=H^2/K_{zz}=\tau_{chem}$. The retrieved P-T profile is used to find the temperature ($T_{\rm quench,C}$) corresponding to the carbon quench pressure ($P_{\rm quench,C}$). Using the posteriors for $P_{\rm quench,C}$ and $T_{\rm quench,C}$, we use equations 12--14 from \cite{zahnle2014} to calculate the chemical reaction timescale \citep[see e.g.][]{xuan2022, deRegt2024}. The pressure scale height $H$ at the quench point is defined as
\begin{equation}
    H = \frac{k_BT_{\rm quench,C}}{\mu mg}
\end{equation}

where $\mu$ is the mean molecular weight of the atmosphere, $m$ is the unified atomic mass, and $g$ is the retrieved surface gravity. From our retrieved parameters, we obtain $\log{K_{zz}} = 4.54^{+0.24}_{-0.23}$ (in units of $\mathrm{cm}^2\mathrm{s}^{-1}$). This value aligns within 3$\sigma$ with the measured diffusion coefficients for other $T_{\rm eff}\sim500\,$K brown dwarfs in literature, such as COCONUTS--2 b ($\log{K_{zz}}\approx$ 3--6; \citealt{2025AJ....169....9Z, kiman2026}), as well as WISE J0430+46 ($\log{K_{zz}}$ = $5.10\pm0.03$) and WISEPC J2056+14 ($\log{K_{zz}}$ = $5.24^{+0.01}_{-0.02}$) from \cite{2024ApJ...976...82T}. However, uncertainties in the mixing length scale and variance of the $K_{zz}$ values with pressure \citep{2022ApJ...938..107M,bieler2024} imply that the $K_{zz}$ obtained from retrievals should be interpreted with caution. The median nitrogen quench pressure $P_{\rm quench,N}>10^{2}$ bars, which is higher than the upper limit of our thermal structure, implying that the NH$_3$ abundance can be vertically constant over our entire P-T profile (Figure \ref{fig:mmrprofile}). Furthermore, as the NH$_3$/N$_2$ abundance ratio contours are nearly parallel to the deep adiabatic thermal profiles of substellar atmospheres, the quenched NH$_3$ abundance is insensitive to the vertical mixing parameter $K_{zz}$ \citep{zahnle2014}.

\subsection{Isotopologue Ratios \label{subsec:isotope}}
Fractionation of carbon isotopes in the protoplanetary disk imply isotopologue ratios can be a tracer of formation location. In particular, planets forming beyond the CO snowline may have a lower $^{12}$CO/$^{13}$CO ratio \cite{2021Natur.595..370Z}. We measure the $^{12}$CO/$^{13}$CO ratio for GJ 504 b, obtaining $^{12}$CO/$^{13}$CO = 72$^{+5}_{-4}$, which agrees excellently with the local ISM value 68 $\pm$ 15 \citep{2005ApJ...634.1126M}. These values are also similar to that of several other directly imaged substellar companions \citep{Gandhi2023, yses2024, xuan2024b, 2025A&A...693A.298G, Gwang2025}. In addition, we also obtain C$^{16}$O/C$^{18}$O = 497$^{+71}_{-61}$, which is consistent with the local ISM value 557 $\pm$ 30 \citep{1999RPPh...62..143W} within $1\sigma$. 


Ices in protoplanetary disks can become enriched in deuterium due to ion-driven pathways at cold temperatures ($<$50 K, \citealt{2014Sci...345.1590C}). Thus enhanced deuterium in substellar atmospheres is a tracer of dominant ice accretion, rather than gas-based accretion (e.g., \citealt{2019ApJ...882L..29M}). This is seen in the envelopes of Uranus and Neptune, which show enhanced deuterium (relative to D/H for the protosolar nebula) due to ice accretion \citep{2013A&A...551A.126F}. This is in contrast to the proto-solar D/H in Jupiter and Saturn, indicating the latter were dominated by gas accretion \citep{2017AJ....154..178P}. While our clear atmosphere retrievals show well-defined posteriors for H$_2$O/HDO and CH$_4$/CH$_3$D ratios (Figure \ref{fig:clearposteriors}), the leave-one-out free retrievals reveal a preference for models without HDO and CH$_3$D opacity (Table \ref{tab:moldetect}). This suggests that well-defined posteriors in the clear atmosphere retrievals could be due to unaccounted-for noise systematics in the science spectrum. Hence, owing to the lack of concrete evidence for presence of HDO and CH$_3$D in GJ 504 b, we do not fit for the H$_2$O/HDO and CH$_4$/CH$_3$D ratios in our cloudy retrievals. 

\subsection{Bolometric luminosity and Effective temperature}
\begin{figure*}
    \centering
    \includegraphics[width=0.6\linewidth]{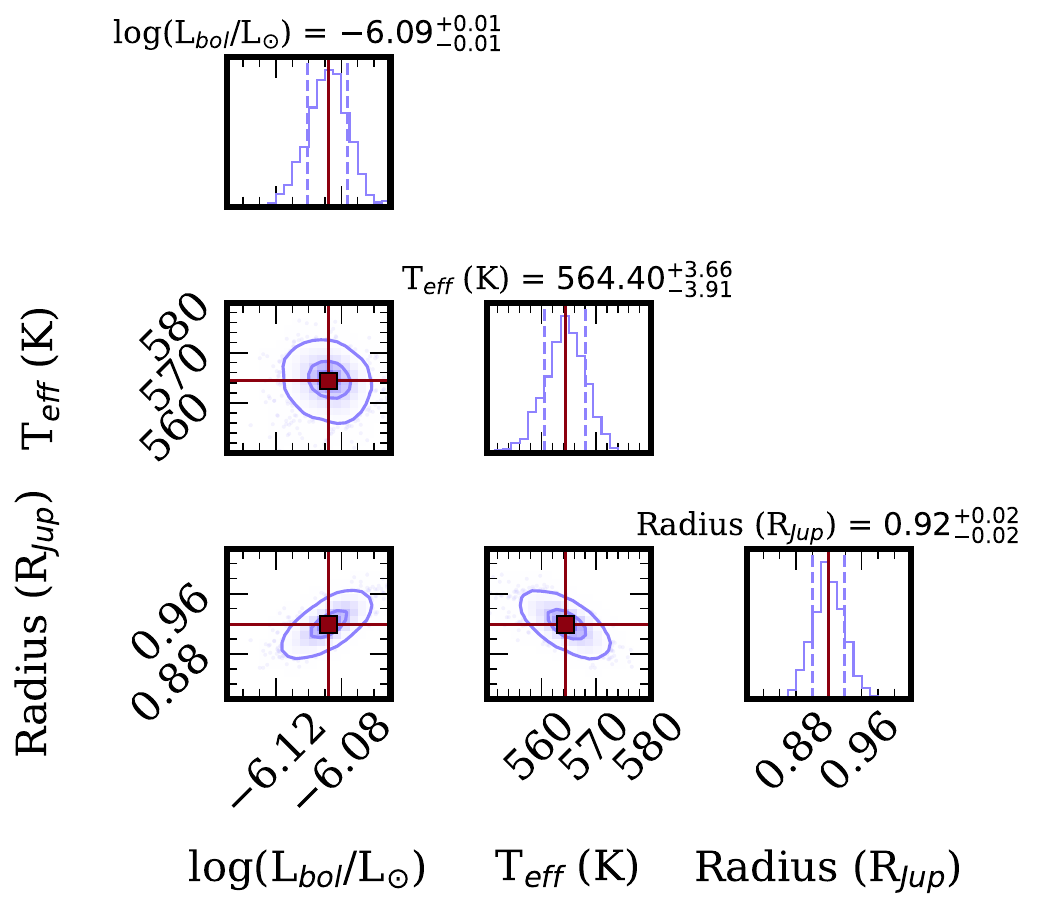}
    \caption{Effective temperature and luminosity of GJ 504 b, obtained using the posterior chains of the cloudy model retrieval with KCl \& ZnS clouds.}
    \label{fig:gj504bteff}
\end{figure*}

Our retrievals involve fitting for the radius of GJ 504 b, giving $R=0.92\pm0.02\,R_\mathrm{Jup}$. We obtain the effective temperate $T_\mathrm{eff}$ using the bolometric luminosity and the companion radius through the  Stefan–Boltzmann law. We use the nested sampling chains generated from the results of our retrieval to generate low-resolution ($R\sim50$) \texttt{petitRADTRANS} models from 0.3--30 $\mu m$, integrating over the entire wavelength range to obtain the bolometric luminosity. These low-resolution models use the same molecular opacity sources previously outlined in Section \ref{subsubsec:opacitysource}. For GJ 504 b, we obtain $\log{L_\mathrm{bol}/L_{\odot}}=-6.09\pm0.01$  which gives an effective temperature $T_\mathrm{eff}=564\pm4$ K (Figure \ref{fig:gj504bteff}).

\subsection{Radial Velocity}
We obtain the first radial velocity measurement for GJ 504 b, though it must be noted that wavelength calibration for the NIRSpec G395H can have systematic uncertainties\footnote{https://jwst-docs.stsci.edu/jwst-calibration-status/nirspec-calibration-status/nirspec-ifu-calibration-status} up to $\sim20\,$kms$^{-1}$. Incorporation of planetary radial velocity as a fitted parameter in our retrievals serves to recalibrate the wavelength solution and ensure that the wavelengths of the models and the data align, rather than provide strong RV constraints from a scientific standpoint. We measure a barycentric-corrected radial velocity measurement $-7.7\pm0.5\,$kms$^{-1}$ for GJ 504 b.

\subsection{Metallicity-Gravity degeneracy \label{subsec:metalvlogg}}
While the current best-fit retrieval models are a good fit to the photometry and NIRSpec spectrum of GJ 504 b (Figure \ref{fig:gj504bestfit}), the posteriors from our atmospheric retrievals indicate that the metallicity and the surface gravity values are correlated (Figure \ref{fig:cloudyposteriors}). Higher gravity makes the spectral features broader and increase flux, which can be suppressed by increasing the metallicity. \cite{skemer2016} also identified these degeneracies in previous modeling work on GJ 504 b. Their work indicates that spectral data around the $K$-band peak at 2.1 $\mu m$ is most effective at resolving this degeneracy due to the contribution of H$_2$ collisional-induced absorption (CIA). Lower H$_2$-CIA opacities in higher metallicity atmospheres lead to higher $K$-band flux, while the dominance of H$_2$-CIA opacity at low metallicities can subdue the $K$-band peak (e.g., \citealt{2012ApJ...750...74S}). On the other hand, surface gravity affects the slope of the $K$-band, rather than the flux of the peak (e.g., \citealt{2013MNRAS.435.2650C}). Regardless of the metallicity values, the retrieved C/O ratio is consistently $\sim0.64$, which is super-solar at $>3\sigma$. Similar metallicity-gravity degeneracy, but with consistently solar C/O was inferred by \cite{2025A&A...693A.298G} in their analysis of $JWST$/NIRSpec data for the brown dwarfs TWA 27A and TWA 28. However, they estimate only marginally super-solar metallicities for both brown dwarfs, as opposed to the highly super-solar metallicities retrieved for GJ 504 b (Table \ref{tab:gj504bmanymodels}).

\subsection{Abundance of nitrogen \label{subsec:nitrogen}}
Carbon, and oxygen consistently show elevated (super-solar) abundances in our retrievals. However, the retrieved nitrogen abundance varies significantly between the various models, and has large measurement errors despite the $10\sigma$ NH$_3$ detection. The retrieved NH$_3$ volume-mixing ratio (VMR) varies between $10^{-4.5}-10^{-4.8}$ depending on the exact cloud prescription used. Considering uncertainties, this is within 2$\sigma$ of the NH$_3$ VMR = $10^{-5.3\pm0.07}$ retrieved by \cite{malin2025} using MIRI photometry.

While NH$_3$ is the only nitrogen-bearing species with strong spectral features in the spectrum of GJ 504 b, other nitrogen-bearing species like N$_2$ and HCN are also present in atmospheres of substellar companions \citep{zahnle2014}. At the super-solar metallicities similar to those of GJ 504 b, \cite{2023ApJ...946...18O} predict that the quenched NH$_3$ abundance is a poor tracer of the bulk nitrogen abundance for substellar objects of all masses and ages, with most of the nitrogen reservoir present in the undetectable N$_2$. Failure to account for N$_2$ results in a low bulk nitrogen abundance, as was estimated by \cite{lew2024} in their analysis of NIRSpec G395H data for the Y-dwarf WISE 1828 ($T_{\rm eff}\sim425-530\,$K). Similarly low bulk nitrogen ([N/H] $<-1$) was also estimated by \cite{meynardie2025} in their analysis of the brown dwarf Ross 458c ($T_{\rm eff}\sim650-770\,$K). 

\cite{2023ApJ...946...18O} describe a prescription to estimate the fraction of the bulk nitrogen (in substellar atmospheres) sequestered in ammonia, which can subsequently be used to calculate the bulk nitrogen abundance N/H. Using their equation 19, we utilize the quenched ammonia abundance, metallicity, and $\log{g}$ from our retrievals, and the effective temperature ($T_{\rm eff}$) to calculate the ratio of the bulk nitrogen abundance and the vertically quenched ammonia abundance ($f_N/f_{NH_3}$). In the case of emission spectroscopy of companions with high interior temperature ($T_{\rm int}\gtrsim300\,$K), the $T_{\rm eff}$ is a good proxy for $T_{\rm int}$ \citep{morley2017,2023ApJ...946...18O}. We then use equation 3 to obtain the bulk N/H. The calculations are performed using retrieved parameters for each of the Na$_2$S, MnS, and KCl + ZnS cloudy models. For the fiducial (KCl \& ZnS) cloud model, we get N/H$=1.7^{+0.9}_{-0.6}$, or [N/H]$=0.22^{+0.19}_{-0.17}$. The Na$_2$S cloud model gives N/H$=4.6^{+10.0}_{-3.0}$, which is equivalent to [N/H]$=0.66^{+0.50}_{-0.46}$. Meanwhile the MnS cloud model gives N/H$=2.0^{+1.2}_{-0.7}$ ([N/H]$=0.30^{+0.20}_{-0.18}$).


However, as discussed in \cite{malin2025}, the quenched NH$_3$ VMR in GJ 504 b depends significantly on the surface gravity of the atmosphere model, hinting at complications in estimating the quenched NH$_3$ abundance (and bulk N/H) due to the metallicity-gravity degeneracy. A detailed chemical kinetics approach (similar to \citealt{Xuan2026}) is needed to obtain an accurate bulk nitrogen abundance which can be used to establish formation. Hence, we just quote the N/H values from our simplistic calculations in Table \ref{fig:enrichmentgj504b}, but do not use them in further discussion on the formation of GJ 504 b.


\section{Discussion \label{sec:discuss}}

\subsection{Comparison to previous photometric analysis}

Previous atmospheric analysis of GJ 504 b involve only ground-based photometry \citep{skemer2016,bonnefoy2018}, or ground-based + $JWST$/MIRI photometry \citep{malin2025}. This work presents the first atmospheric analysis and modeling of GJ 504 b using spectroscopic data, in addition to previous photometry. Joint atmospheric modeling of NIRSpec/G395H spectroscopy, in addition to previous photometry, gives $T_\mathrm{eff}=564\pm4$ K, $\log{L_{bol}/L_{\odot}}=-6.09\pm0.01$, R$=0.92\pm0.02\,R_\mathrm{Jup}$, $\log{g}=4.87^{+0.13}_{-0.12}$, and $[C/H]=[M/H]=0.67^{+0.13}_{-0.12}$ for our best-fitting cloudy model (KCl+ZnS clouds; Table \ref{tab:gj504bmanymodels}). 

We first compare with \cite{skemer2016}, where the luminosity, radius, and [M/H] all agree within $1\sigma$. \cite{bonnefoy2018} consider only the luminosity and temperature from their atmospheric analysis, obtaining other parameters (radius, gravity) from various isochronal analyses. Their $L_{bol}$ and $T_\mathrm{eff}$ agree well with our values. Lastly, comparisons with \cite{malin2025} reveal excellent agreement with their luminosity, [M/H] and C/O. However, the radius from their NIR+MIR analyses are discrepant from our retrievals. In general, we notice a trend where photometry-only forward modeling analyses, as well as isochronal analyses assuming a young age for the system ($<0.5$ Gyr) give lower temperatures and gravities. However, isochronal analyses assuming an old age for the system ($>2.0$ Gyr) agree excellently with our best retrieved parameters (Table \ref{tab:litcompare}). Meanwhile, the order of magnitude differences in surface gravity compared to \cite{malin2025} are primarily explained by the differences in data. As they discuss in their work, photometric points rely only on the shape of the continuum for surface gravity measurements, whereas more accurate measurements of surface gravity are obtained from the shape of the individual lines (e.g., \citealt{2004ApJ...600.1020M,2013ApJ...772...79A,2017ApJ...838...73M}). In our work, we are able to resolve the NH$_3$ spectral lines (especially at $3.2\,\mu m$, refer Figure \ref{fig:gj504bestfit}), which can be used as a proxy for surface gravity \citep{zahnle2014}. However, we cannot exclude the possibility that modeling assumptions in our retrievals as compared to self-consistent models might be responsible for the discrepancies. Regardless, the best spectral features for measurements of surface gravity are present in the $J$, $H$, $K$-bands (Na I, K I, H$_2$ collision-induced absorption; e.g., \citealt{2007ApJ...667..537L,2013ApJ...772...79A}), making spectral information at those wavelengths critical for better constraints on the surface gravity. 

\begin{deluxetable*}{c|cccccccc}
    \centering
    \tabletypesize{\footnotesize}
    \tablecaption{Comparison to atmospheric parameters of GJ 504 b from literature \label{tab:litcompare}}
    \tablewidth{0pt}
    \tablehead{\colhead{Work} & \colhead{$T_\mathrm{eff}$ (K)} & \colhead{$\log{L_{bol}/L_{\odot}}$} & \colhead{Radius ($R_\mathrm{Jup}$)} & \colhead{$\log{g}$} & \colhead{[M/H]} & \colhead{C/O Ratio} & \colhead{Model/s used} & \colhead{Comments}}
    \startdata
    \CellWithForceBreak{This work \\ (KCl + ZnS clouds)} & $564\pm4$ & $-6.09\pm0.01$ & $0.92\pm0.02$ & $4.87^{+0.13}_{-0.12}$ & $0.67^{+0.13}_{-0.12}$ & $0.64\pm0.02$ & \texttt{petitRADTRANS} & \CellWithForceBreak{G395H spectra \\ + photometry} \\ \hline
    \cite{kuzuhara2013}\tablenotemark{a} & $510^{+30}_{-20}$ & $-6.09^{+0.06}_{-0.08}$ &  & $3.9^{+0.4}_{-0.2}$ & & & {1} & \CellWithForceBreak{From evo. \\ models} \\ \hline
    \cite{skemer2016}\tablenotemark{b} & $544\pm10$ & $-6.13\pm0.03$ & $0.96\pm0.07$ &  & $0.60\pm0.12$ & & \CellWithForceBreak{Custom \\ model grid} & \CellWithForceBreak{Forward \\ modeling} \\ \hline
    \multirow{2}{*}{\cite{bonnefoy2018}}\tablenotemark{a} & $550\pm50$ & $-6.15\pm0.15$ &  &  &  & & \CellWithForceBreak{2--6} & \CellWithForceBreak{Forward \\ modeling} \\ \cline{2-9}
    & $537^{+68}_{-64}$ & $-6.11\pm0.18$ & $0.95^{+0.08}_{-0.06}$& $4.82^{+0.19}_{-0.27}$ & & & {1,7} & \CellWithForceBreak{From evo. models }\\
    \hline
    \cite{malin2025} & $512\pm10$ & $-6.12\pm0.02$ & $1.08^{+0.04}_{-0.03}$ & $3.45^{+0.35}_{-0.25}$ & $0.54^{+0.09}_{-0.11}$ & 0.70$^{+0.06}_{-0.07}$ & {4} & \CellWithForceBreak{NIR+MIR \\ photometry}
    \enddata
\tablerefs{\texttt{petitRADTRANS} - \cite{molli2019,molli2020}, (1) \cite{baraffe2003}, (2) \cite{skemer2016}, (3) \cite{2013MSAIS..24..128A} (\texttt{BT-SETTL}), (4) \cite{baudino2015,2018ApJ...854..172C} (\texttt{Exo-REM}), (5) \cite{2015ApJ...813...47M} (\texttt{petitCODE}), (6) \cite{2017ApJ...842..118L} (\texttt{ATMO}), (7) \cite{2008ApJ...689.1327S}}
\tablenotetext{a}{Evolutionary models comparisons from \cite{kuzuhara2013} are assuming a young system (0.1--0.5 Gyr), while comparisons from \cite{bonnefoy2018} are assuming an old system ($4.0\pm1.8$ Gyr).}
\tablenotetext{b}{\cite{skemer2016} used a custom model grid based on previous work by \cite{2012ApJ...756..172M,2014ApJ...787...78M} to fit for atmospheric parameters of GJ 504 b. This grid was also subsequently used by \cite{bonnefoy2018} in their analysis.} 

\end{deluxetable*}

\subsection{Age and Mass of GJ 504 b \label{subsec:agemass}}

\begin{figure*}
    \centering
    \plottwo{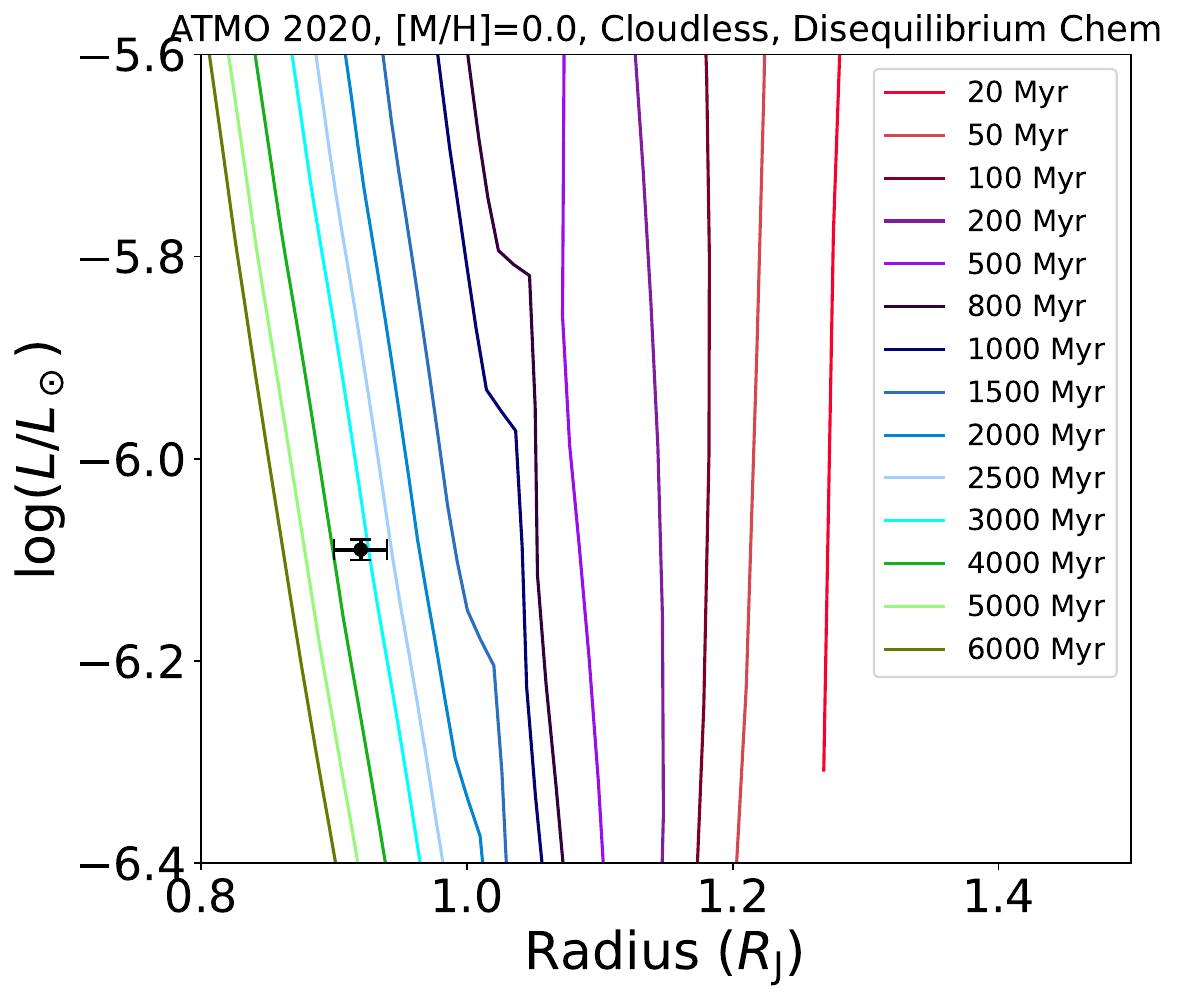}{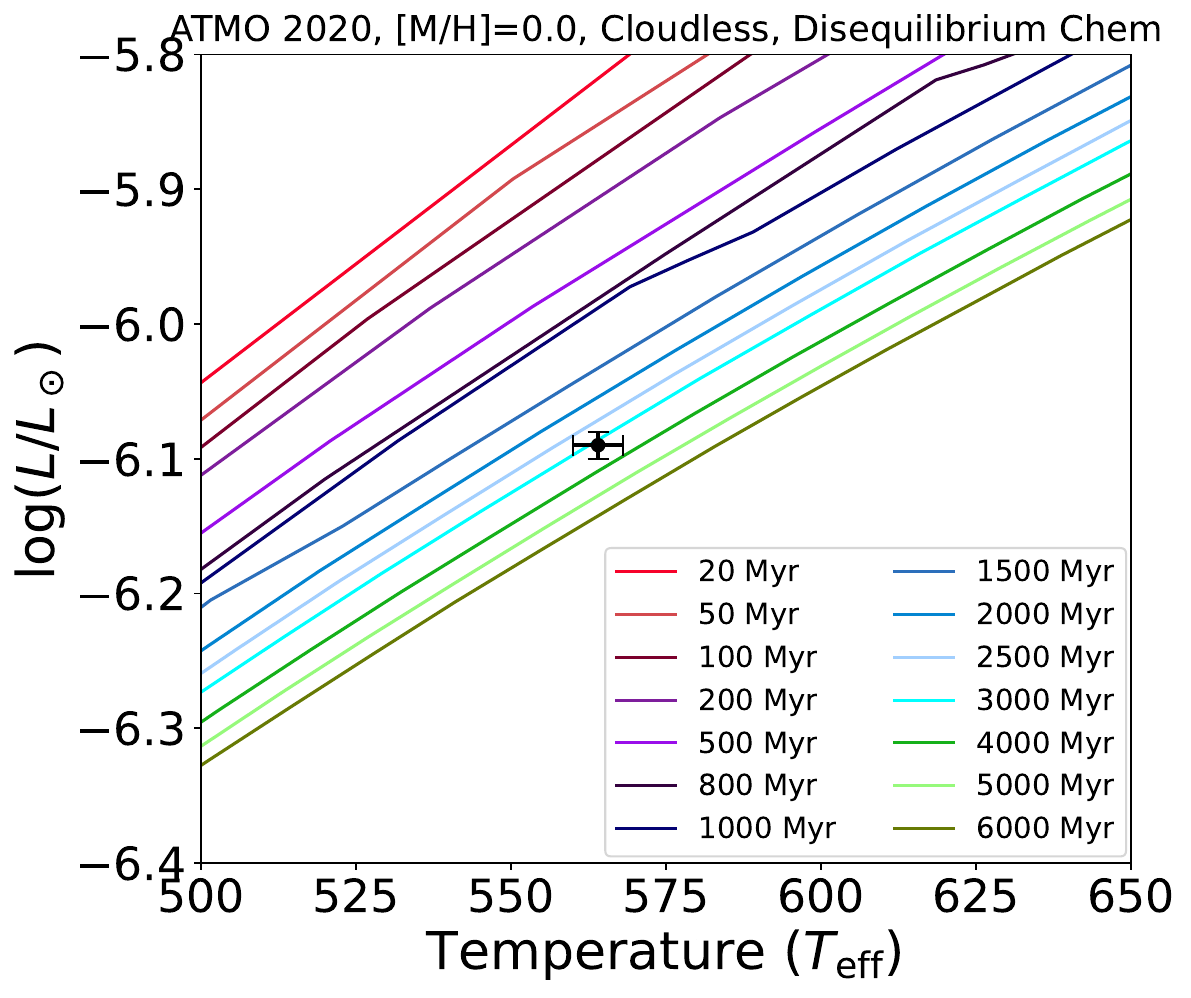}
    \caption{ATMO 2020 isochrones \citep{2020A&A...637A..38P} for solar metallicity, cloudless atmospheres with disequilibrium chemistry. The isochrones show the dependence of luminosity and radius \textit{(left)}, and luminosity and temperature \textit{(right)} on age of a substellar object. GJ 504 b is marked with a solid black circle, with its retrieved luminosity, temperature, and radius indicating an age 2.5--4.0 Gyr.}
    \label{fig:atmo2020}
\end{figure*}

Our retrieval analysis fits for the gravity and radius of GJ 504 b. Using the retrieved gravity and radius from the best-fit model (KCl+ZnS clouds), we obtain a mass of 25.2$^{+8.4}_{-6.0}$ $M_\mathrm{Jup}$. Across all cloud models (Table \ref{tab:gj504bmanymodels}), we obtain masses in the 20--30 $M_\mathrm{Jup}$ range. This result agrees excellently with the masses for a 4.0 $\pm$ 1.8 Gyr companion obtained by \cite{bonnefoy2018} (23$^{+10}_{-9}$ $M_\mathrm{Jup}$) and \cite{2015ApJ...806..163F} ($\sim25\,M_\mathrm{Jup}$). We also compare the retrieved luminosity, temperature, and radius to the ATMO 2020 evolutionary tracks \citep{2020A&A...637A..38P} and those from \cite{2008ApJ...689.1327S}. The ATMO tracks are generated using solar metallicity, cloudless models with both equilibrium and disequilibrium chemistry prescriptions. We use the tracks with disequilibrium chemistry to obtain a companion age of 2.5--4.0 Gyr (Fig. \ref{fig:atmo2020}; similar approach to \citealt{malin2025}). This estimate is in good agreement with the system age of 2.11 $\pm$ 0.46 Gyr from \cite{2025A&A...694A.179P}. However, it is possible that the retrieved radii are simply underestimated, with a $2\sigma$ discrepancy compared to radii predicted by the 2 Gyr ATMO isochrones at similar luminosities ($\sim0.96-0.97\,R_{\rm Jup}$). The retrieved bulk parameters correspond to ATMO evolutionary mass around 19--27 $M_\mathrm{Jup}$.

\begin{figure*}
    \centering
    \plottwo{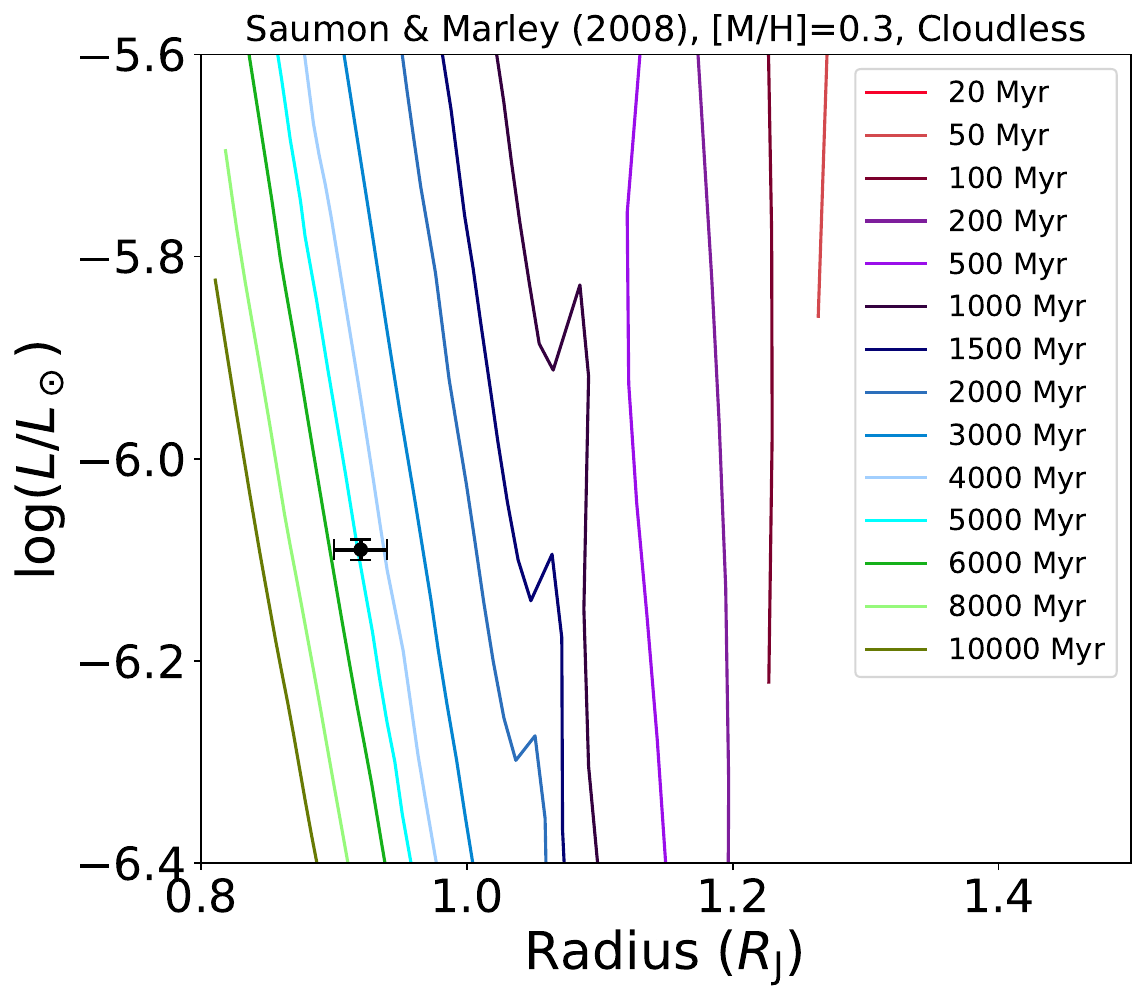}{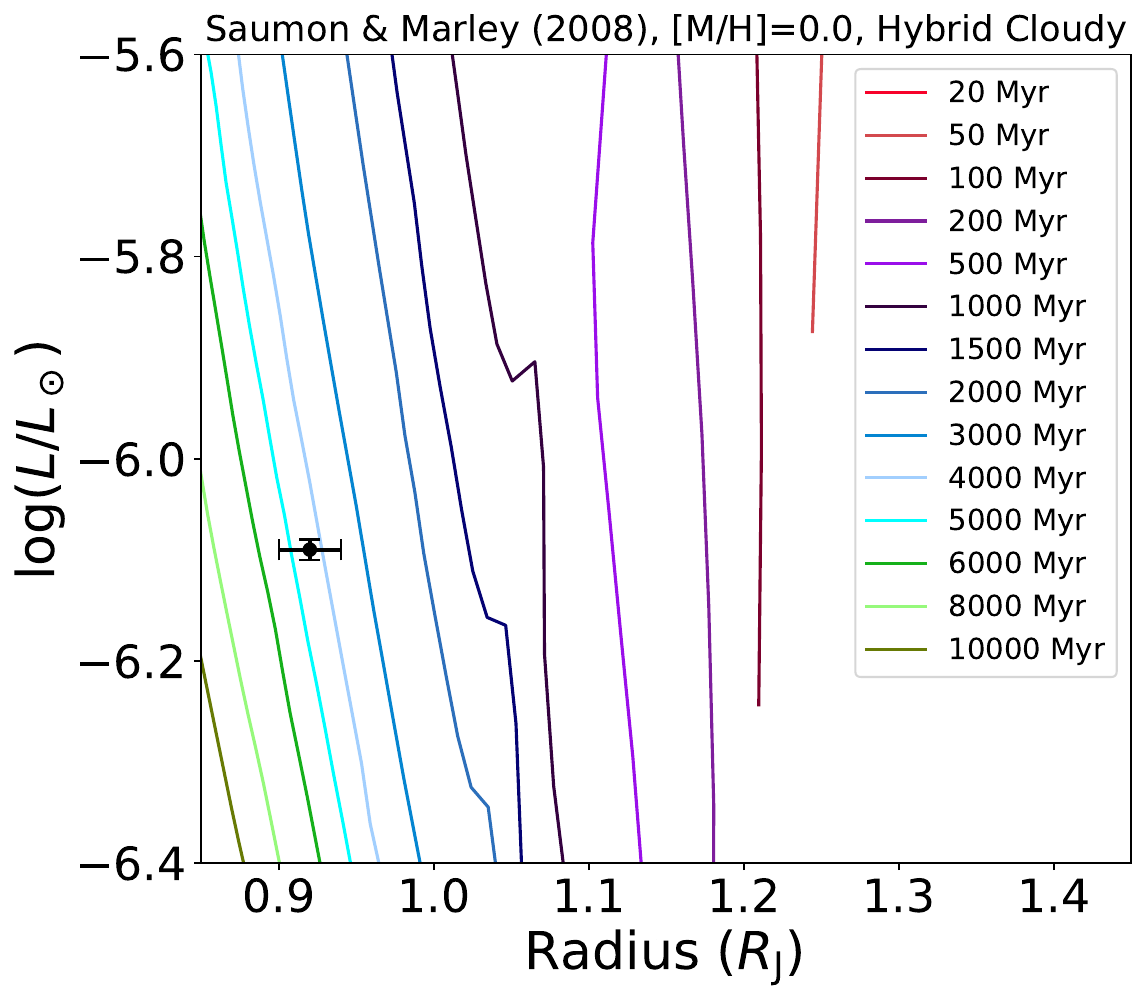}
    \caption{Isochrones from \cite{2008ApJ...689.1327S} showing the dependence of luminosity and radius on age of a substellar object. \textit{(Left)} Evolutionary models corresponding to high-metallicity, cloudless atmospheres, and \textit{(Right)} Evolutionary models corresponding to solar metallicity, hybrid cloudy/cloudless atmospheres. GJ 504 b is marked with a solid black circle.}
    \label{fig:sm2008}
\end{figure*}

Meanwhile, the models by \cite{2008ApJ...689.1327S} have a range of different prescriptions, including sub-solar to super-solar metallicities, and both cloudy and clear atmospheres. Comparisons to these evolutionary models give an age of 4.0--6.0 Gyr for high-metallicity ([M/H]$\,=0.3$), cloudless atmospheres and 4.0--5.0 Gyr for solar metallicity, hybrid cloudy/cloudless atmospheres (Fig. \ref{fig:sm2008}). For both sets of models, this corresponds to masses around 24--30 $M_\mathrm{Jup}$. The evolutionary masses obtained using both ATMO 2020 and \cite{2008ApJ...689.1327S} tracks agree excellently with the mass for a $\sim4\,$Gyr substellar companion (23$^{+10}_{-9}$ $M_\mathrm{Jup}$) from \cite{bonnefoy2018}. More recently, \cite{malin2025} performed a forward-modeling analysis of near-infrared and mid-infrared photometry of GJ 504 b to determine the atmospheric properties of GJ 504 b, followed by comparison of the luminosity, temperature, and radius to the ATMO evolutionary models with disequilibrium chemistry \citep{2020A&A...637A..38P}. Their best-fit radius ($1.08^{+0.04}_{-0.03}\,R_\mathrm{Jup}$) corresponds to model ages 400 Myr -- 1 Gyr, which is equivalent to masses from $1$--$17\,M_\mathrm{Jup}$ for GJ 504 b. While their masses agree with our values at just over 1$\sigma$, there is tension in the radius and age estimates.

Overall, our investigations indicate that GJ 504 b is an 2.5--6.0 Gyr planetary-mass companion with super-stellar metallicity. However, the above comparisons also underscore a lack of evolutionary models that account for high metal enrichment ([M/H]$\,>0.5$), cloudy atmospheres, and disequilibrium chemistry, especially for late-T (and colder) planetary-mass companions. Studies of GJ 504 b over a broader wavelength range will enable benchmarking of upcoming atmospheric/evolutionary models especially generated to understand cold, enriched gas giants and planetary-mass companions that might be discovered with $JWST$.

\subsection{Metal enrichment of GJ 504 b and Implications for Formation}

Atmospheric metal enrichment of GJ 504 b has been observed by previous studies exclusively analyzing photometric data \citep{skemer2016,malin2025}. In this work, we confirm the results of prior studies, consistently obtaining super-solar metallicities using a variety of model prescriptions. However, accurate interpretation of atmospheric abundances for PMCs in the context of their formation requires comparisons with corresponding host star abundances (e.g., \citealt{aneesh2025a,aneesh2025b}).

\subsubsection{Elemental Abundances of GJ 504 A}
As a solar-type star within 20 pc, multiple studies have investigated the chemical profile of GJ 504 A. In general, previous work have determined super-solar metallicities [M/H]$\,=$0.1--0.3 for the star (e.g., \citealt{1993A&A...275..101E, 2004A&A...418..551M, 2005ApJS..159..141V,2007PASJ...59..335T,2010MNRAS.403.1368G,2012A&A...541A..40M,2012A&A...542A..84D,2013ApJ...764...78R, dorazi2017, 2017AJ....153...21L, 2018A&A...614A..55A,2021AJ....161..134H,aneesh2025a}). Of these, \cite{dorazi2017} and \cite{aneesh2025a} also determined individual metal abundances for carbon (C), oxygen (O), and sulfur (S). The former used FEROS spectra from the ESO archives to obtain abundances of 14 elements, including carbon (C) and oxygen (O). They find [C/H]$\,\sim$[O/H]$\,\sim$[S/H]$\,\sim0$ (i.e., solar) using an equivalent width analysis. Meanwhile, \cite{aneesh2025a} measured abundances of 15 elements using Lick/APF-Levy spectra, finding [C/H] and [O/H] enhanced by $\sim0.3\,dex$. For comparison with the abundances of GJ 504 b, we adopt [C/H] and [O/H] measurements from the latter study, as they measure the abundances for the two elements using both spectral fitting and equivalent width methods, leading to greater accuracy of their results. In particular, we adopt their spectral fit [C/H]$\,=0.27\pm0.03$ and [O/H]$\,=0.28\pm0.11$ measurements, owing to their better precisions. For consistency and minimization of systematics, we also adopt their [S/H]$\,=0.42\pm0.17$. The only nitrogen abundance for GJ 504 A is from \cite{1985ApJ...289..556L}, who obtain [N/H]$\,=0.14\pm0.20$.

\begin{deluxetable*}{c|c|ccccccc}
\centering
\tabletypesize{\normalsize}
\tablecaption{Metal Abundances of GJ 504 A and GJ 504 b\label{tab:hostcompare}}
\tablewidth{0pt}
\tablehead{\colhead{} & \colhead{Model} & \colhead{Carbon} & \colhead{Oxygen} & \colhead{Nitrogen$^{1}$} & \colhead{Sulfur}}
\startdata
\hline \hline
GJ 504 A &  & & & & \\ \hline
[X/H]    & - & $0.27\pm0.03$ & $0.28\pm0.11$ & $0.14\pm0.20$ \tablenotemark{\small{2}} & $0.42\pm0.17$ \\ \hline \hline
GJ 504 b &  & & & & \\ \hline
[X/H]    & \multirow{2}{*}{EddySed cloud -- KCl+ZnS} & $0.67^{+0.13}_{-0.12}$ & $0.60^{+0.13}_{-0.11}$ & $0.22^{+0.19}_{-0.17}$ & $0.59^{+0.15}_{-0.14}$ \\
X/H\tablenotemark{\small{3}}  &   & $2.5^{+0.9}_{-0.6}$ & $2.1^{+1.0}_{-0.6}$ & $1.2^{+1.1}_{-0.5}$ & $1.5^{+1.0}_{-0.5}$ \\ \hline
\enddata
\tablenotetext{1}{As discussed in Section \ref{subsec:nitrogen}, we only mention the nitrogen abundance but do not use the values for formation discussions.} \tablenotetext{2}{GJ 504 A [N/H] from \cite{1985ApJ...289..556L}} \tablenotetext{3}{X/H$\,>1$ indicates GJ 504 b is enriched in the specific metal relative to GJ 504 A. X/H$\,=1$ implies no enrichment.}
\end{deluxetable*}

\subsubsection{Formation of GJ 504 b from Metal Enrichment Profile}

\begin{figure*}
    \centering
    \plottwo{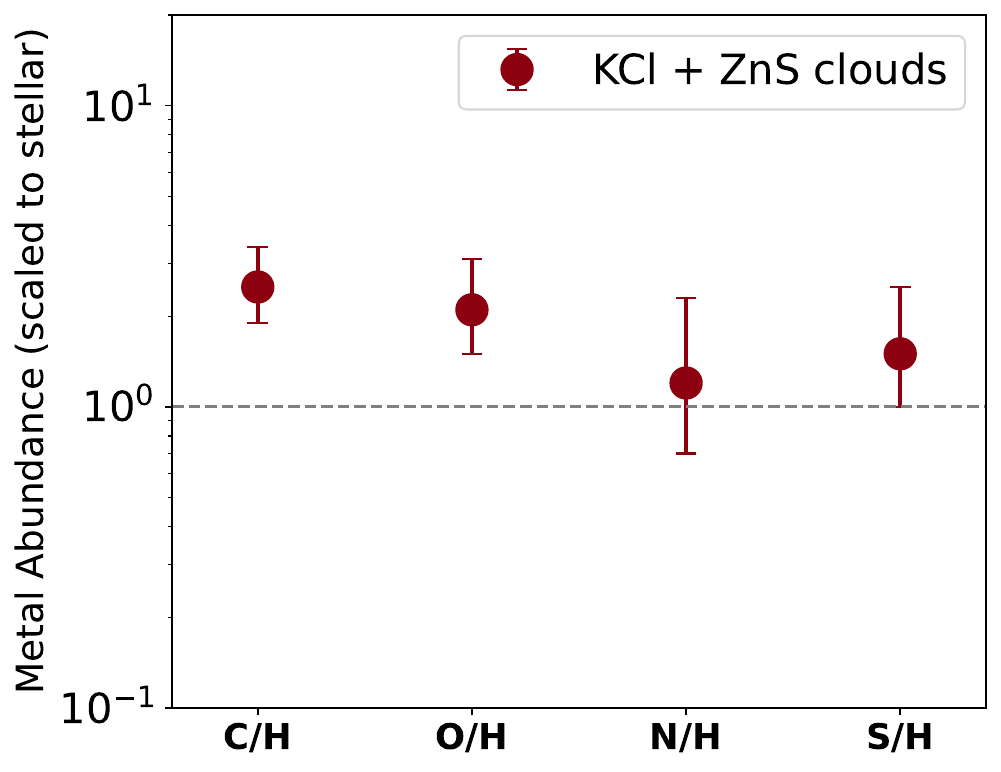}{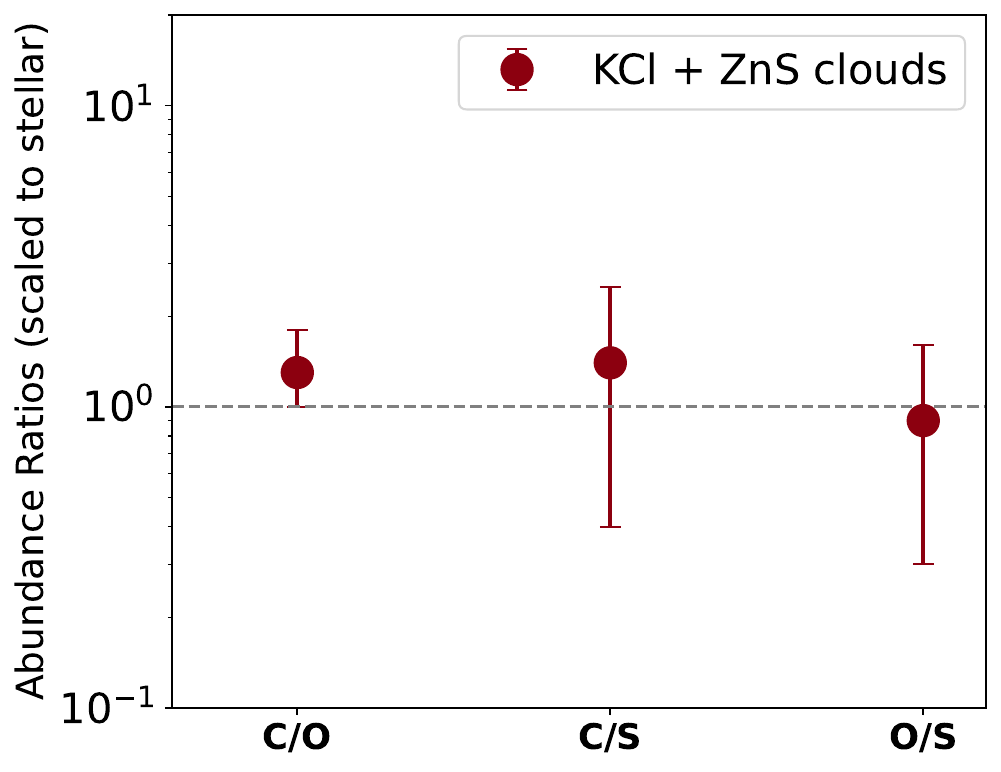}
    \caption{(\textit{Left}) Metal enrichment of GJ 504 b relative to stellar (dashed line). C is enriched 2.5$\times$ stellar, O has an enrichment of 2.1$\times$ stellar. N is stellar, though we heavily caveat this measurement (Section \ref{subsec:nitrogen}). The S abundance is stellar, indicating a lack of atmospheric enrichment by solid accretion. (\textit{Right}) Abundance ratios for GJ 504 b, with the dashed line indicating the stellar values. All three abundance ratios are stellar within 1$\sigma$. Both figures use the abundance values provided in Table \ref{tab:hostcompare}.}
    \label{fig:enrichmentgj504b}
\end{figure*}

We use the metal abundances of GJ 504 b and its primary GJ 504 A (refer Table \ref{tab:hostcompare}) to estimate the metal enrichment profile of the companion, depicted in Figure \ref{fig:enrichmentgj504b}. From our fiducial (KCl + ZnS) cloud model, GJ 504 b is enriched in C and (tentatively) O relative to the primary, with C/H $=2.5\times$ stellar and O/H $=2\times$ stellar. Meanwhile, S/H is stellar within $1\sigma$. We also use the NH$_3$ detection to infer stellar N/H$=1.2^{+1.1}_{-0.5}$, though we heavily caveat this measurement (refer Section \ref{subsec:nitrogen}). Overall our retrieved metal enrichment profile for GJ 504 b reveals all metals being stellar within at most 2.5$\sigma$.

Recent literature has suggested atmospheric metal enrichment as a better diagnostic of planet-like formation \citep{jiwang2025}, with the revised giant planet mass-metallicity relations by \cite{2025ApJ...994...43C} providing considerable evidence for significant solid accretion even during the runaway gas accretion phase of giant planet formation. Along these lines, our metal enrichment only tentatively supports a planetary nature for GJ 504 b. However, there are substantial uncertainties even regarding the abundances of the primary, with elevated carbon, oxygen, sulfur measurements by \cite{aneesh2025a}, but solar values from \cite{dorazi2017}. In case of the latter, the observed abundances of GJ 504 b would indicate metal enrichment in all three of carbon, oxygen, and sulfur. Hence, we briefly discuss scenarios which could lead to metal enrichment for this companion.

Sulfur is expected to be mostly found in the solid form through most of the protoplanetary disk \citep{2019ApJ...885..114K}, making atmospheric enrichment of sulfur an indicator of solid accretion \citep{2023ApJ...952L..18C,Xuan2026}. A lack of atmospheric sulfur enrichment, but accompanied by atmospheric enrichment in carbon and oxygen (as seen here) would indicate a lack of solid accretion, with metal enrichment occurring primarily through accretion of metal-enriched gas \citep{sb2021a,sb2021b}. However, the large uncertainties on the planetary sulfur abundance, as well as the sulfur abundance of the primary, imply that any conclusions regarding the extent of solid accretion are still tentative.

Metal enrichment through metal-rich gas occurs as the inward drifting pebbles lose their volatile content via evaporation at the snowlines, enriching the local disk gas and subsequently, the PMC atmosphere. Similar levels of enrichment for C and O, imply formation interior to the CO snowline. 
Resolution of the metallicity-gravity degeneracy would enable stronger constraints on the individual metal abundances, allowing us to test whether any potential enrichment of this PMC was mostly by gas accretion, or if significant solid accretion was also involved.

We seek to explore these abundances in the context of the protoplanetary disk around the natal GJ 504 A and location of the various snowlines. We use the equation for the thermal structure of the disk around a star of mass $M_*$ and luminosity $L_*$ from \cite{Chiang1997, 2016A&A...591A..72I}:
\begin{equation}
\begin{split}
T_{disk} \simeq 150\,{\rm K} \, \times \, (M_*/M_{\odot})^{-1/7} \,(L_*/L_{\odot})^{2/7} \\ (r/{\rm 1\,au})^{-3/7}
\end{split}
\end{equation}
Substituting $M_*=1.29\,M_{\odot}$ for GJ 504 A \citep{dorazi2017,2025A&A...694A.179P} and assuming that $L_* = L_{\odot}(M_*/M_{\odot})^{3/2}$ for a pre-main sequence star \citep{Choi2016, Chachan2023b}, we obtain:
\begin{equation}
T_{disk} \simeq 160 \, \times (r/{\rm 1\,au})^{-3/7} \, \rm K
\end{equation}
This thermal structure is appropriate for a 1--5 Myr protoplanetary disk. The CO snowline ($\sim 30$ K) would be at a distance of $\sim$50 au for the disk around GJ 504 A. The current separation of GJ 504 b is $\sim$43.5 au \citep{kuzuhara2013}, a distance just inside the CO snowline. Our inference from GJ 504 b's composition that it accreted CO-rich gas interior to the CO snowline is therefore in concordance with its current location and does not require it to have undergone large scale migration. We note that there is some uncertainty in the sublimation temperatures and snowline locations of CO (at the level of $\sim 5$ K, e.g., \citealt{Fayolle2016, 2016ApJ...833..203P}), which in turn might affect the inferred extent of migration. Regardless, GJ 504 b's current location is compatible with region in which we expect it to have accreted CO rich gas.


Recent work on the HR 8799 planets has revealed super-stellar abundances of C, O, S for HR 8799 cde \citep{Ruffio2026, Xuan2026} and C, O, S, and N for HR 8799 b \citep{Xuan2026}. All four planets show 3--5$\times$ enrichment in carbon and oxygen relative to their host star. Such enrichment levels are similar to Jupiter in our Solar System, which is $\sim3\times$ enriched in carbon, oxygen, nitrogen, and sulfur relative to the Sun \citep{2004Icar..171..153W,2020NatAs...4..609L}. Our analysis indicates GJ 504 b could have possible metal enrichment, with carbon and oxygen enriched 2--2.5$\times$ relative to the primary GJ 504 A. Furthermore, the mass-metallicity relations from \cite{2025ApJ...994...43C} reveal that planetary bulk metallicities plateau at super-stellar values rather than decreasing with increasing planetary mass, even for super-Jupiters ($M_p>2\,M_{\rm Jup}$). These findings indicate that super-Jupiters preferentially accrete heavy metals even during their runaway gas accretion phase, leading to enhanced atmospheric metallicities. This highlights the importance of atmospheric metallicity, rather than planet mass, as the distinguishing criteria between planets and brown dwarfs. 

Overall, the large uncertainties on its metal enrichment prevent us from definitively concluding that GJ 504 b formed liked a planet, with a brown dwarf nature (stellar abundances) not entirely discounted. Future work with the MIRI MRS data (GO 3647) or measurements of its dynamical mass for resolving the metallicity-gravity degeneracy are needed to solve the mystery regarding the formation of GJ 504 b. In addition, our work highlights the importance of accurate host star abundance measurements while inferring the planetary nature of substellar companions.

\section{Conclusions \label{sec:conclusions}}
We present the analysis of medium-resolution ($R\sim2,700$), 2.9--5.3 $\mu m$ NIRSpec observations of the planetary-mass companion GJ 504 b as part of the $JWST$ program GTO 2778. This companion has previously proved too faint for ground-based spectroscopic observations.

$\bullet$ Utilizing the \texttt{BREADS} forward modeling (FM) framework, we detect GJ 504 b at a combined S/N = 357 across the entire F290LP wavelength range. This framework was also used to obtain a spline-filtered planet spectrum, which is used for subsequent atmospheric modeling. \\ \\

$\bullet$ We demonstrate the first successful PSF subtraction using Angular Differential Imaging (ADI) in the NIRSpec point cloud, detecting GJ 504 b at a S/N~=~14.7 in NRS1 and S/N~=~28.8 in NRS2. ADI-based speckle subtraction is able to reach contrasts of $10^{-4}$ at 1$^{\prime\prime}$ and fainter than $10^{-5}$ at 2.5$^{\prime\prime}$. However, the NIRSpec undersampling and current point-cloud linear interpolation scheme implies that the approach is speckle-noise dominated. Regardless, ADI successfully detects GJ 504 b and provides continuum information FM cannot. An observing strategy that incorporates both approaches would enable recovery of the PMC continuum information and constraints on atmospheric parameters (effective temperature, radius), while also enabling the detection and abundance measurements for trace molecules. \\ \\

$\bullet$ The extracted spectrum reveals strong signatures of molecular species like CO$_2$ ($82\sigma$), NH$_3$ ($10\sigma$), $^{13}$CO ($36\sigma$), C$^{18}$O ($12\sigma$), and H$_2$S ($6\sigma$). The multitude of species enable strong constraints on the metallicity, and the C/O ratio. \\ \\
$\bullet$ Joint retrieval analysis of the G395H spectrum and previous photometry provides measurements of the carbon, oxygen, nitrogen, and sulfur abundances. Additionally, our models reveal a strong preference for cloudy atmospheres over clear (cloud-free) atmospheres, with clouds especially needed to prevent isothermality in the thermal structure. \\ \\

$\bullet$ The retrieved radius and gravity for the best-fit cloudy model provide a mass 25.2$^{+8.4}_{-6.0}$ $M_\mathrm{Jup}$ for GJ 504 b. Comparisons of the retrieved luminosity, radius, and temperature with various evolutionary model grids indicate that GJ 504 is an old (2.5--6.0 Gyr) system. Better evolutionary models for low-temperature companions ($T_{\rm eff}\,<$550 K) which also incorporate effects of super-solar metallicities and clouds are needed to accurately capture the age of this companion. \\ \\

$\bullet$ We find a $2.5^{+0.9}_{-0.6}\times$ enhancement in carbon, a $2.1^{+1.0}_{-0.6}\times$ enhancement in oxygen and stellar sulfur relative to the primary GJ 504 A. The observed metal enrichment profile provides tentative evidence for planet-like formation for GJ 504 b, but does not completely exclude a brown dwarf nature. Future work on more accurate stellar abundances or reducing uncertainties in the enrichment values (e.g., with dynamical mass measurements) is needed to break the degeneracies in the formation and nature of GJ 504 b. 

\section*{acknowledgements}
This paper reports work carried out in the
context of the JWST Telescope Scientist Team  (\href{https://www.stsci.edu/∼marel/jwsttelsciteam.html}{https://www.stsci.edu/$\sim$marel/jwsttelsciteam.html}; PI: M. Mountain). Funding is provided to the team by NASA through grant 80NSSC20K0586. Based on observations with the NASA/ESA/CSA JWST, associated with program GTO-2778 (PI: Marshall Perrin), obtained at the Space Telescope Science Institute, which is operated by AURA, Inc., under NASA contract NAS 5-03127.  The data described here may be obtained from the MAST archive at
\dataset[doi:10.17909/yavj-1e54]{https://doi.org/10.17909/yavj-1e54}.
A.B. and Q.M.K acknowledge support by the National Aeronautics and Space Administration under Grants/Contracts/Agreements No. 80NSSC24K0210 issued through the Astrophysics Division of the Science Mission Directorate. Any opinions, findings, and conclusions or recommendations expressed in this work are those of the author(s) and do not necessarily reflect the views of the National Aeronautics and Space Administration. J.W.X is grateful for support from the Heising-Simons Foundation 51 Pegasi b Fellowship (grant \#2025-5887).

This work used the Anvil supercomputer at the Purdue Rosen Center for Advanced Computing \citep{10.1145/3491418.3530766} through allocation PHY250125 from the Advanced Cyberinfrastructure Coordination Ecosystem: Services \& Support (ACCESS) program, which is supported by National Science Foundation grants \#2138259, \#2138286, \#2138307, \#2137603, and \#2138296. 

This research has made use of the SVO Filter Profile Service ``Carlos Rodrigo" (\href{https://svo2.cab.inta-csic.es/theory/fps/}{https://svo2.cab.inta-csic.es/theory/fps/}), funded by MCIN/AEI/10.13039/501100011033/ through grant PID2023-146210NB-I00.

This work was conducted at the University of California, San Diego, which was built on the unceded territory of the Kumeyaay Nation, whose people continue to maintain their political sovereignty and cultural traditions as vital members of the San Diego community.

\facilities{JWST (NIRSpec) \citep{jakobsen2022, boker2022}, MAST \citep{2018SPIE10704E..13M}}
\software{SciPy \citep{2020SciPy-NMeth}, NumPy \citep{harris2020array}, matplotlib \citep{Hunter:2007}, astropy \citep{astropy:2013, astropy:2018, astropy:2022}, corner \citep{corner}, STPSF \citep{2012SPIE.8442E..3DP,2014SPIE.9143E..3XP}, petitRADTRANS \citep{molli2019,molli2020}, jwst \citep{bushouse2024}, whereistheplanet\footnote{https://whereistheplanet.com/}, h5py\footnote{https://docs.h5py.org/}, pandas \citep{reback2023pandas}, astroquery \citep{2019AJ....157...98G}, BREADS \citep{agrawal2024}, species \citep{species}, dynesty \citep{speagle2020}, easyCHEM \citep{2024arXiv241021364L}}

\bibliographystyle{aasjournal}
\bibliography{main}

\appendix
\restartappendixnumbering

\section{Noise covariance in the high-pass filtered spectrum \label{appendix:hpfacf}}
\begin{figure*}[h!]
    \centering
    \includegraphics[width=1.0\linewidth]{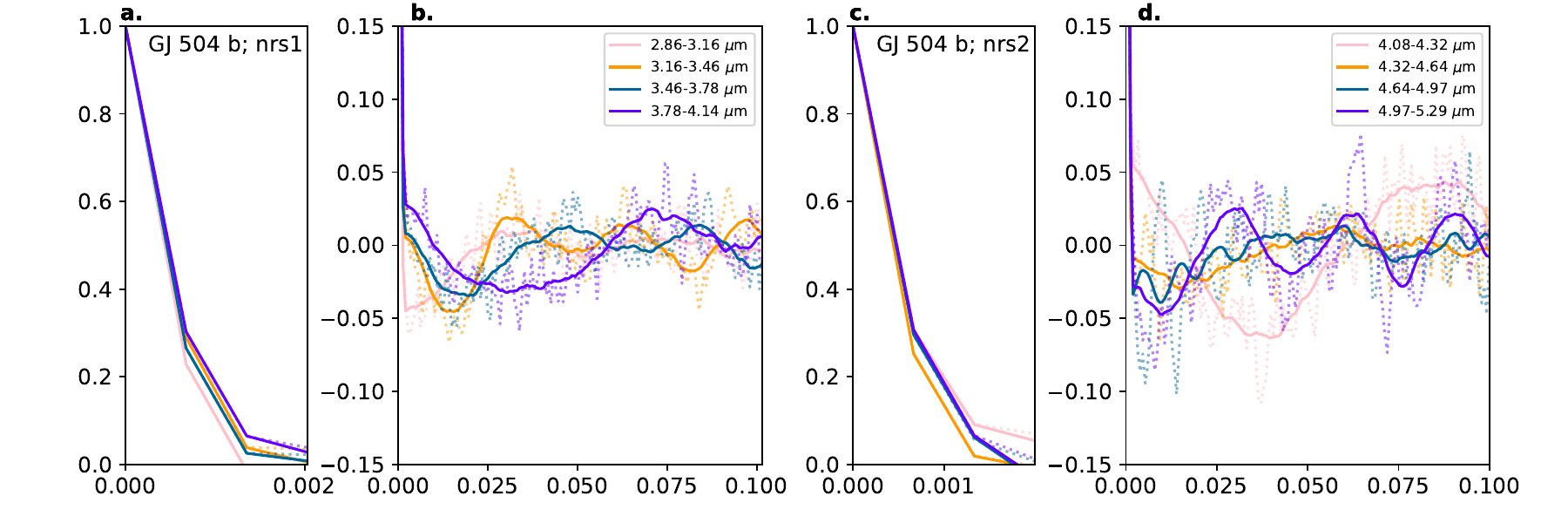}
    \caption{Autocorrelation profile of the noise for the companion spectrum. \textbf{a.} shows the small-scale correlation in the noise in the NRS1 spectrum due to white-noise and interpolation of the detector images onto a regular wavelength grid. \textbf{b.} shows the large-scale correlations in the noise due to the residual speckles and limited number of spline nodes. \textbf{c.} and \textbf{d.} are the same plots but for NRS2.}
    \label{fig:acfhpf}
\end{figure*}

The noise covariance for the high-pass filtered spectrum is computed along similar lines to \cite{Ruffio2026}. The covariance matrix is defined as a 4$\times$4 block diagonal matrix to account for the variable node spacing of the spline nodes, with the off-diagonal blocks set to null. The autocorrelation of the noise (smoothed using a 10-pixel sliding-window mean filter) for each section of the spectrum is used to derive the four block correlation matrices. The covariance matrices are obtained by scaling the correlation matrices such that the diagonal elements correspond to the flux errors on the companion spectrum. The empirical correlation profile for the high-pass filtered spectrum is shown in Figure \ref{fig:acfhpf}.

\section{Noise covariance in the ADI spectrum \label{appendix:adiacf}}
\begin{figure}[H]
\centering
\includegraphics[trim={3cm 0 3cm 0},clip, width=1.0\linewidth]{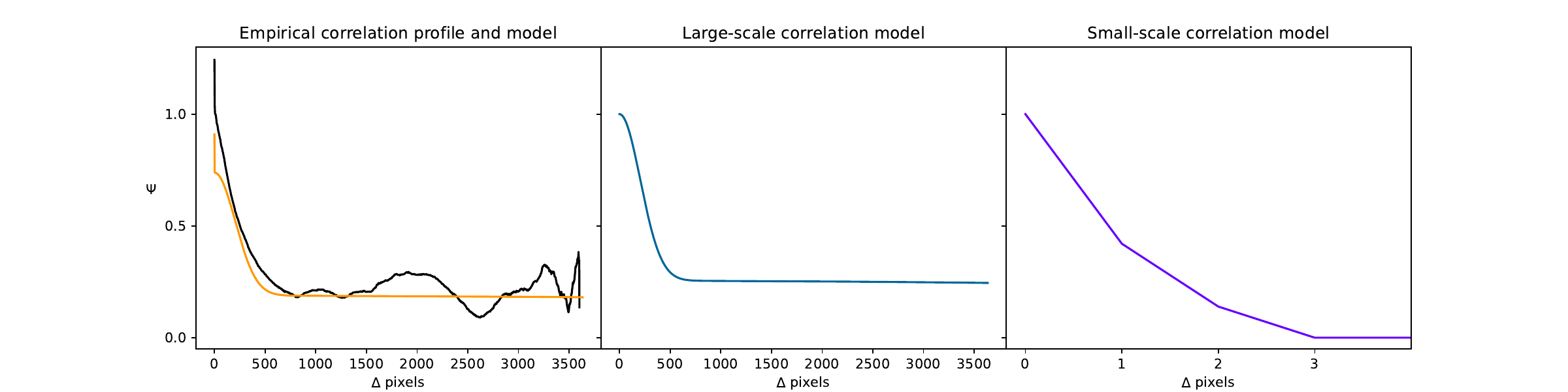}
\caption{(\textit{Left}) The empirical noise correlation profile of the ADI spectrum of GJ 504 b (black) and the best-fit model (orange) to the same. (\textit{Middle}) The model for the correlated portion of the empirical profile. (\textit{Right}) The uncorrelated portion of the empirical profile arising due to interpolation effects and affecting only the first two pixels (note the different x-axis).}
\label{fig:adicov}
\end{figure}

\restartappendixnumbering
The error budget for ADI for the NIRSpec IFU is dominated by the systematics from the spatial interpolations involved in fitting the ``reference" point cloud generated using one of the observation rolls to the other roll. As mentioned previously, the $JWST$ wavefront is extremely stable and is unlikely to be the cause of these imperfect subtractions. In addition, these uncertainties on the ADI spectrum are correlated. We adopt the approach discussed in Appendix E of \cite{ruffio2024} to obtain the semi-empirical covariance matrix for the ADI spectrum of GJ 504 b. The empirical correlation profile for the ADI spectrum and the models used to derive our covariance matrix are shown in Figure \ref{fig:adicov}. 

\section{Posterior plots for atmospheric retrievals}
\restartappendixnumbering
In this section we present the posterior plots for the clear retrieval and the cloudy retrieval with KCl \& ZnS clouds.

\begin{figure}
\centering
\includegraphics[trim={0cm 0 0cm 0},clip, width=1.0\linewidth]{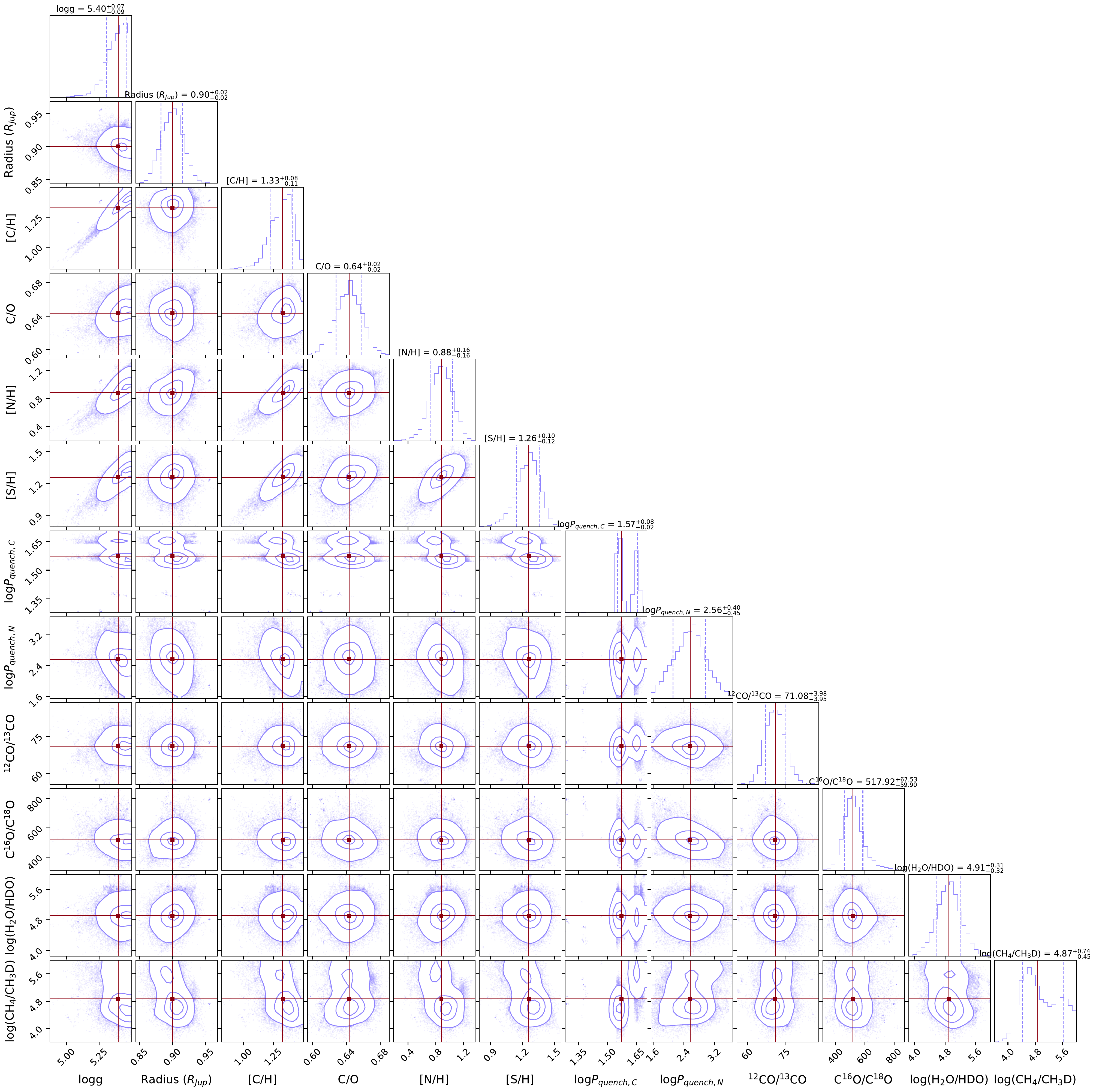}
\caption{Posterior plot for the \texttt{petitRADTRANS} clear retrieval. The 1-D marginalized posteriors are shown along the diagonal. Red lines represent the 50 percentile with the blue dotted lines representing the 16 and 84 percentiles. The subsequent covariances between all the parameters are in the corresponding 2-D histograms.}
\label{fig:clearposteriors}
\end{figure}

\begin{figure}
\centering
\includegraphics[trim={0cm 0 0cm 0},clip, width=1.0\linewidth]{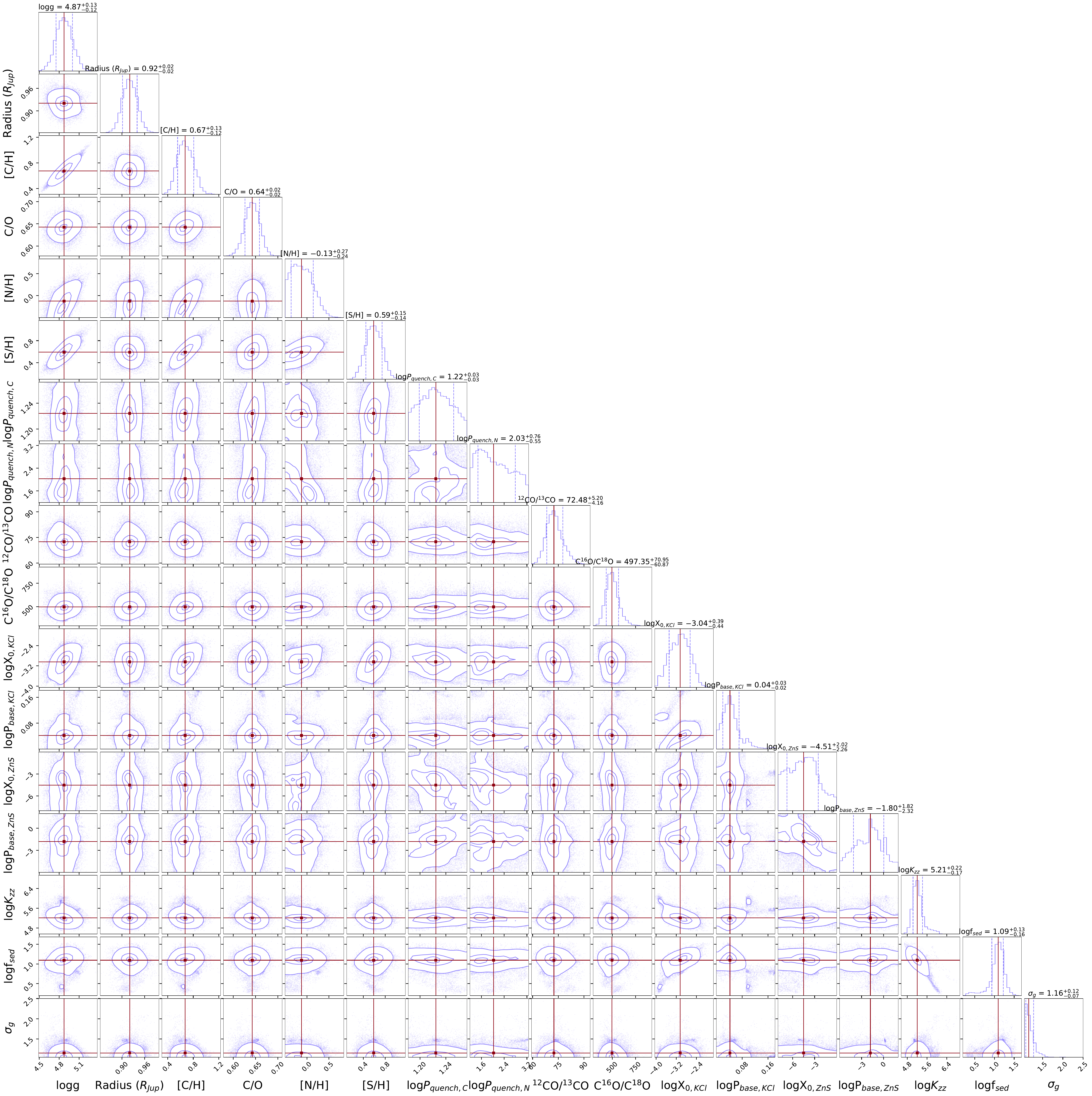}
\caption{Posterior plot for the \texttt{petitRADTRANS} cloudy retrieval with KCl \& ZnS clouds. The 1-D marginalized posteriors are shown along the diagonal. Red lines represent the 50 percentile with the blue dotted lines representing the 16 and 84 percentiles. The subsequent covariances between all the parameters are in the corresponding 2-D histograms.}
\label{fig:cloudyposteriors}
\end{figure}

\end{document}